\documentclass[useAMS,usenatbib]{mn2e}
\usepackage[]{txfonts}
\usepackage{subfigure}
\usepackage{epsfig}
\graphicspath{{graphs/}{../ps/}{ps/}}
\usepackage{url}
\usepackage[normalem]{ulem}

\def\simless{\mathbin{\lower 3pt\hbox
{$\rlap{\raise 5pt\hbox{$\char'074$}}\mathchar"7218$}}}   
\def\simmore{\mathbin{\lower 3pt\hbox
{$\rlap{\raise 5pt\hbox{$\char'076$}}\mathchar"7218$}}}   
\newcommand       \apj          {ApJ}
\newcommand       \apjl         {ApJL}
\newcommand       \aap          {A\&A}
\newcommand       \nat          {Nature}
\newcommand       \mnras        {MNRAS}

\newcommand       \aj      {AJ}
\newcommand       \prd      {Phys.~Rev.~D.~}
\newcommand       \prc      {Phys.~Rev.~C.~}

\newcommand       \araa      {ARA\&A}

\newcommand       \pasj   {PASJ}
\newcommand      \apjs {ApJ Supplements}

\newcommand      \physrep {Physics Reports}
\newcommand      \jcap {JCAP}
\newcommand \ssr {Space Science Reviews}
\newcommand \actaa {Acta Astronomica}

\newcommand{\be}{\begin{eqnarray}}
\newcommand{\ee}{\end{eqnarray}}

\newcommand{\De}{\Delta}
\newcommand{\eps}{\varepsilon}

\def\gtrsim{\mathrel{\hbox{\rlap{\hbox{\lower4pt\hbox{$\sim$}}}\hbox{$>$}}}}

\title[Neutrino-heated winds from proto-magnetars]{Neutrino-heated winds from millisecond proto-magnetars as sources of the weak $r$-process}

\author[Vlasov et al.]{Andrey D.~Vlasov$^{1}\thanks{E-mail: adv2110@columbia.edu}$, Brian D.~Metzger$^{1}$, Jonas Lippuner$^{2}$, Luke F. Roberts$^{3}$,
\newauthor
 Todd A.~Thompson$^{4}$\\
$^{1}$Department of Physics and Columbia Astrophysics Laboratory, Columbia University, New York, NY, 10027, USA\\ $^{2}$TAPIR, Walter Burke Institute for Theoretical Physics, Mailcode 350-17, California Institute of Technology, Pasadena, CA 91125, USA\\ $^{3}$NSCL and Department of Physics and Astronomy, Michigan State University, East Lansing, MI 48824\\ $^{4}$Department of Astronomy and Center for Cosmology and Astro-Particle Physics, The Ohio State University, Columbus, OH 43210, USA}

\begin{document}
\twocolumn
\date{Received / Accepted}
\pagerange{\pageref{firstpage}--\pageref{lastpage}} \pubyear{2014}

\maketitle

\label{firstpage}


\begin{abstract}
We explore heavy element nucleosynthesis in neutrino-driven winds from rapidly-rotating, strongly magnetized proto-neutron stars (``millisecond proto-magnetars") for which the magnetic dipole is aligned with the rotation axis, and the field is assumed to be a static force-free configuration.  We process the proto-magnetar wind trajectories calculated by \citet{Vlasov+14} through the $r$-process nuclear reaction network SkyNet using contemporary models for the evolution of the wind electron fraction during the proto-neutron star cooling phase.  Although we do not find a successful second or third peak $r$-process for any rotation period $P$, we show that proto-magnetars with $P \sim 1-5$ ms produce heavy element abundance distributions that extend to higher nuclear mass number than from otherwise equivalent spherical winds (with the mass fractions of some elements enhanced by factors of $\gtrsim 100-1000$).  The heaviest elements are synthesized by outflows emerging along flux tubes which graze the closed zone and pass near the equatorial plane outside the light cylinder.  Due to dependence of the nucleosynthesis pattern on the magnetic field strength and rotation rate of the proto-neutron star, natural variations in these quantities between core collapse events could contribute to the observed diversity of the abundances of weak $r$-process nuclei in metal-poor stars.  Further diversity, including possibly even a successful third-peak $r$-process, could be achieved for misaligned rotators with non-zero magnetic inclination with respect to the rotation axis.  If proto-magnetars are central engines for GRBs, their relativistic jets should contain a high mass fraction of heavy nuclei of characteristic mass number $\bar{A}\approx 100$, providing a possible source for ultra-high energy cosmic rays comprised of heavy nuclei with an energy spectrum that extends beyond the nominal GZK cut-off for protons or iron nuclei.

\end{abstract}

\section{Introduction} 
\label{sec:intro}
The astrophysical sites responsible for synthesizing the heaviest elements in the Universe via the rapid neutron capture process ($r$-process, \citealt{Burbidge+57,Cameron57}) have been debated for decades (for reviews, see \citealt{Qian&Wasserburg07,Arnould+07,Sneden+08,Thielemann+11}). 

Core collapse supernovae (SNe) have long been considered potential $r$-process sites.  This is in part due to their short delays following star formation, which allows the earliest generations of metal-poor stars in our Galaxy (e.g.~\citealt{Mathews+92}, \citealt{Sneden+08}), or satellite dwarf galaxies (e.g.~\citealt{Roederer16}), to be polluted with $r$-process elements prior to significant iron enrichment.  Throughout the 1990s, the high entropy neutrino-heated winds from proto-neutron stars (PNS) (\citealt{Duncan+86,Qian&Woosley96}), which emerge on a timescale of seconds after a successful explosion as the PNS deleptonizes, were considered the most likely $r$-process site\footnote{
Another $r$-process mechanism in the core collapse environment results from $\nu-$induced spallation in the He shell (e.g., \citealt{Banerjee+11}).  This channel is limited to very low metallicity $Z \lesssim 10^{-3}$ and thus cannot represent the dominant $r$-process source over the age of the galaxy, though it could be important for the first generations of stars.  } within the core collapse environment (e.g., \citealt{Meyer+92,Woosley+94}).  A high entropy, or correspondingly low density, results in an $\alpha$-rich freeze-out of the 3- and effective 4-body reactions responsible for forming seed nuclei in the wind (\citealt{Woosley&Hoffman92}).  The resulting higher ratio of neutrons to seed nuclei then allows neutron captures to proceed to heavier elements than if the protons were instead entirely trapped in heavy seeds.  

However, some contemporary (e.g., \citealt{ Takahashi+94}) and many subsequent calculations of the wind properties (e.g., \citealt{Qian&Woosley96,Kajino+00,Sumiyoshi+00,Otsuki+00,Thompson+01, Arcones+07, Roberts+10, MartinezPinedo+12, Roberts+12b, Fischer+12}) showed that the requisite combination of low electron fraction $Y_e \lesssim  0.5$ and high entropy needed to reach the second or third $r$-process peaks (\citealt{Hoffman+97}) were unlikely to be reached.  Possible exceptions include a very massive PNS (\citealt{Cardall&Fuller97}), or given the presence of additional wind heating by damping of convectively-excited acoustic or Alfv\'en waves (\citealt{Suzuki+Nagataki05,Metzger+07}) or of non-standard physics, such as an eV-mass sterile neutrino (e.g., \citealt{Tamborra+12,Wu+14}). 

If all SNe produced the $r$-process in equal quantities, the required mass of $r$-process elements per event to explain Galactic abundances is relatively low, $\sim 10^{-5}M_{\odot}$  (e.g.~\citealt{Macias&Ramirez-Ruiz16}).  However, several lines of evidence instead support much `higher yield' $r$-process events being common in our Galaxy, both now and in its early history.  These include the detection of $^{244}$Pu on the ocean floor at abundances roughly 2 orders lower than that expected if the source were frequent, low-yield events like those predicted from PNS winds in normal SNe \citep{Wallner+15,Hotokezaka+15}.  A fraction of the stars in the dwarf galaxy Reticulum II are highly enriched in $r$-process elements, indicating that this galaxy was polluted early in its history by a single $r$-process event with a yield much higher than the standard neutrino-driven wind \citep{Ji+16}.  
Based on an analysis of the mass swept up by the SN blast wave, \citet{Macias&Ramirez-Ruiz16} argue that, in order to explain the largest enhancements in [Eu/Fe] at low metallicites, individual $r$-process events must synthesize at least $10^{-3.5}M_{\odot}$ of $r$-process material, again significantly higher than predicted by standard PNS wind models.  Such high abundances, if present in young Galactic SN remnants, would be detectable by their radioactive gamma-ray or X-ray decay lines (\citealt{Qian+98,Ripley+14}).

Increasingly, the mergers of compact binary neutron star systems are seen as promising alternative sites for at least the heaviest $r$-process elements (\citealt{Lattimer&Schramm74,Eichler+89,Freiburghaus+99}).  General relativistic hydrodynamical simulations of the merger events show that some of the matter ejected dynamically during the merger retains a sufficiently low electron fraction ($Y_e \lesssim 0.1$) to form the heavy $r$-process elements extending beyond the third $r$-process peak at atomic number $Z \gtrsim 78$ (e.g., \citealt{Goriely+11,Wanajo+14}).  

Less clear in the merger framework is the origin of the charged particle process nuclei\footnote{These elements are sometimes also referred to as light element primary process nuclei (\citealt{Arcones&Montes11}).}  ($Z = 38-40$) and of the `light' or `weak' $r$-process nuclei ($Z \approx 41-55$) nuclei.  It is well known that $r$-process nuclei with $Z \lesssim 56$ in metal-poor stars show greater star-to-star variation in their abundance patterns than the heavier $r$-process nuclei (e.g.~\citealt{Honda+06,Roederer+10}), despite showing an apparently robust pattern similar to the solar abundances in the heavier range $56 \lesssim Z \lesssim 76$ (e.g.~\citealt{Sneden+08}).  Such abundances variations are again difficult to explain from standard PNS winds, which, other than due to usually modest variations in the PNS mass, should produce broadly similar yields for each supernova.  This has motivated considering alternative sources for the light $r$-process nuclei in the environments of neutron star mergers, such as accretion disk winds (\citealt{Fernandez&Metzger13,Perego+14,Just+15,Martin+15b,Wu+16}) or components of the dynamical ejecta with higher electron fractions (\citealt{Wanajo+14,Goriely+15}). 

This paper focuses on another variation on the canonical picture of neutrino-driven wind that occurs if the PNS is formed rapidly-rotating, with an ultra-strong magnetic field $B \gtrsim 10^{14}-10^{15}$ G, similar to those of Galactic magnetars (a so-called ``millisecond proto-magnetar"; \citealt{Thompson03}).  If magnetic fields are dynamically-important during the SN explosion itself, magnetic acceleration of matter away from the PNS can act to lower the asymptotic electron fraction of the unbound ejecta by preventing the outflowing matter from coming into equilibrium with neutrino absorption reactions (\citealt{Metzger+08b}), possibly facilitating a heavy $r$-process \citep{Metzger+08a,Winteler+12,Nishimura+15,Nishimura+16}.  Although such models have great potential, quantitative studies of MHD-SNe are still in their infancy (e.g.~\citealt{Takiwaki+16}).  High resolution three-dimensional simulations are needed to resolve the dynamo responsible for tapping into the shear kinetic energy to generate a large-scale magnetic field (\citealt{Guilet&Muller15,Mosta+15,Sawai&Yamada16}).  They are also needed to capture the growth of non-axisymmetric (e.g.~magnetic kink and sausage mode) instabilities, which can disrupt MHD jet-like structures that otherwise are stable in axisymmetric simulations (e.g., \citealt{Mosta+14}).  The observed rate of presumably MHD-powered, hyper-energetic SNe is also low compared to the total core collapse rate (e.g.~\citealt{Podsiadlowski+04}), thus requiring a high $r$-process yield per event to explain the Galactic abundances through this channel alone.  

Even if rotation and magnetic fields are not dynamically important during the explosion phase itself, they will become so during the subsequent neutrino-wind phase (\citealt{Thompson03,Thompson+04,Metzger+07}).  Indeed, such a situation is potentially much more common than a full-blown MHD-powered SN, because the  angular momentum of the progenitor stellar core needed to produce a PNS rotating with a period of $P \gtrsim 3-4$ ms is probably lower than that needed to explode the star (although see \citealt{Thompson+05}).  Furthermore, the impact of such a moderately rapidly-rotating magnetar on the properties of the explosion, such as its total energy and $^{56}$Ni yield, could be comparatively modest (e.g.~\citealt{Suwa&Tominaga15}).  The total birth rate of Galactic magnetars is estimated to exceed 10\% of the total core collapse rate (\citealt{Woods&Thompson06}), but the birth rotation rates are largely unconstrained observationally.  
A small population of core collapse supernovae with extremely high optical luminosities may result from millisecond magnetar birth (e.g.~\citealt{Kasen&Bildsten10,Woosley10,Metzger+14}).  However, the limited range of rotation periods and surface dipole magnetic field strengths which result in greatly enhanced SN emission imply that only a fraction of all magnetar births manifests this way.  

\citet[hereafter V14]{Vlasov+14} solved for the steady-state structure of neutrino-heated winds from rotating, magnetized PNS, building on the previous one-dimensional equatorial monopole calculations of \citet{Metzger+07}.  V14 assumed that the magnetic field structure was that of an axisymmetric aligned dipole under the force-free approximation \citep{Timokhin06}.  This is valid provided that the energy density of the magnetic field greatly exceeds that of gas pressure and the kinetic energy density, as they find to be valid throughout the radii where the most important nucleosynthesis occurs for surface dipole field strengths of $B_{\rm d} \gtrsim 10^{14}-10^{15}$ G, depending on the neutrino luminosity (V14, their Fig.~3).  V14 obtained a series of one-dimensional solutions calculated along flux tubes corresponding to different polar field lines.  They stitched together these flux tube solutions to determine the global wind properties across the entire open magnetosphere at a fixed neutrino luminosity and rotation period.  

V14 found that proto-magnetars with rotation periods of $P \sim 2-3$ ms produce outflows more favorable for the production of third-peak $r$-process nuclei.  This is due to their much shorter expansion times $\tau_{\rm dyn}$ through the seed nucleus formation region, yet only moderately lower entropies $s$, as compared to spherical non-rotating PNS winds.  They found that the critical ratio of $s^{3}/\tau_{\rm dyn}$ was higher than for spherical winds, but not sufficiently so that nucleosynthesis would proceed to the third $r$-process peak at $A \sim 200$, based on the analytic criteria of \citet{Hoffman+97}.  Except in the case of extremely rapid rotation, near the centrifugal break-up period of $P \sim 1$ ms, magnetic acceleration was not found to significantly enhance the mass outflow rate per unit surface area.  In fact, the total mass loss rate $\dot{M}$ is in most cases substantially lower than in an otherwise equivalent spherical wind because outflows occur only along open magnetic field lines, which thread only a small fraction ($\lesssim 10 \%$, depending on the rotation period) of the total PNS surface.

The present paper explores the nucleosynthetic yield of magnetar birth in greater detail by processing the wind trajectories from V14 through the nuclear reaction network SkyNet (\citealt{Lippuner&Roberts15}) in order to determine the detailed wind abundance pattern ($\S\ref{sec:methods}$).  Even in the non-rotating spherical case, several critical wind properties remain uncertain; most notably the electron fraction $Y_e$ depends sensitively on the differences between the luminosities and mean energies of the electron neutrinos and anti-neutrinos diffusing out from the PNS interior.  The accuracy of these luminosities and neutrino energies is currently limited by uncertainties in the radiation transport and microphysics.  Thus, in most cases we focus on comparing our calculated yields to those of otherwise equivalent spherical winds, in order to isolate the diversity in the nucleosynthesis imprinted exclusively by magneto-rotational effects  ($\S\ref{sec:results}$). Although we find that proto-magnetars are probably incapable of synthesizing the heaviest $r$-process elements, at least during the neutrino wind phase, their winds may still contribute to  the inferred Galactic diversity of weak $r$-process sources ($\S\ref{sec:weak}$).

Beyond their role as possible sources of $r$-process nucleosynthesis and engines for powering luminous supernovae, millisecond magnetars are contenders for the central engines powering gamma-ray bursts (\citealt{Usov92,Wheeler+00,Thompson+04,Metzger+11a}a). If GRBs are indeed powered by the rotational energy of a magnetar, then the nucleosynthesis products of their winds will be directly entrained in the relativistic jet which escapes from the star and powers the prompt gamma-ray emission.  As we describe in $\S\ref{sec:uhecr}$, this unique jet composition could have implications for the composition of ultra-high energy cosmic rays if they are accelerated in GRB jets (\citealt{Metzger+11b}b).

\begin{figure}
\begin{center}
\includegraphics[width=1.0\linewidth]{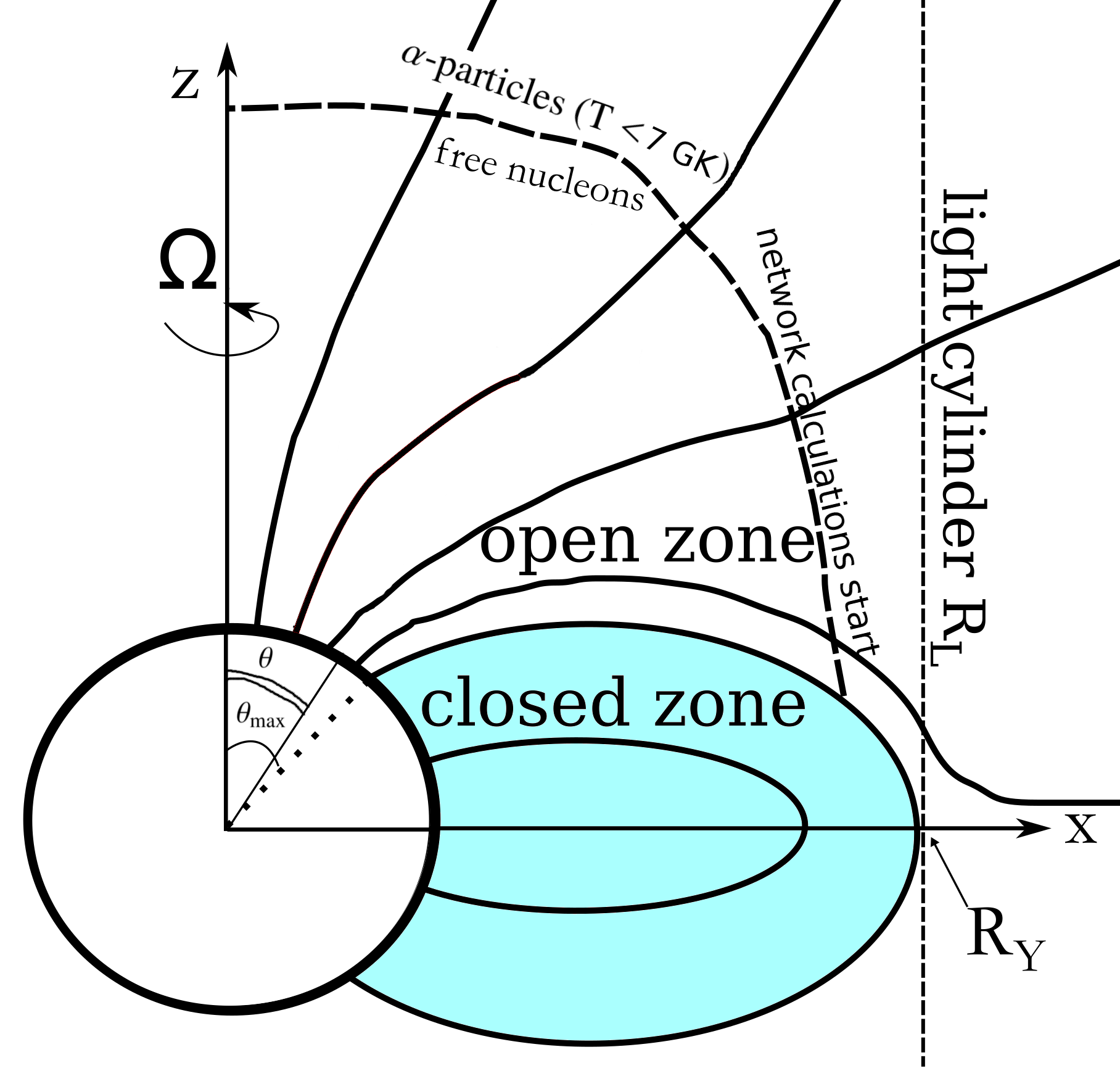}
\caption{Geometry of neutrino-heated winds from magnetized aligned rotating proto-neutron stars.  Outflows occur from the `open zone',  along field lines with polar angles $\theta < \theta_{\rm max}$, where $\theta_{\rm max} = \sin^{-1}\left[R_{\rm ns}/R_{\rm Y}\right]$ is the angle of the last closed field line, $R_{\rm ns}$ is the PNS radius, $R_{\rm Y} \lesssim R_{\rm L}$ is the radius at which the last closed field line crosses the equator (the `Y point'), $R_{\rm L} = c/\Omega$ is the light cylinder radius, and $\Omega$ is the angular rotation rate of the PNS.  In the illustration and in our calculations we have assumed that $R_{\rm Y} = R_{\rm L}$. The network calculations start once the nucleons in the wind partially recombine to $\alpha-$particles, at $T=7\cdot 10^9$ K.}
\label{fig:schematic}
\end{center}
\end{figure}

\section{Nuclear Reaction Network Calculations}
\label{sec:methods}

\subsection{Thermodynamic Trajectories}
\label{sec:traj}

In the limit of force-free electrodynamics, the geometric structure of the proto-magnetar wind from aligned rotator is fully specified by the magnetar rotation period $P$ and the Y point radius $R_{\rm Y}$, which defines the intersection between the last open field lines of the polar cap and the equatorial plane.  We operate under the assumption that $R_{\rm Y}$ equals the radius of the light cylinder, $R_{\rm L} = c/\Omega = 2\pi c P$, where $\Omega = 2\pi/P$ is the magnetar rotational angular frequency (Fig. \ref{fig:schematic}).  Unbound outflows occur along open magnetic flux bundles, with the outflow geometry varying with latitude $\theta$ from the pole at $\theta =0$ to the last open field line at $\theta = \theta_{\rm max} \simeq \sin^{-1}\left[R_{\rm ns}/R_{\rm L}\right]$, where $R_{\rm ns} = 12$ km is the assumed NS radius.  

Our nucleosynthesis calculations are performed along steady-state, one-dimensional wind trajectories for different field lines $\theta$, as calculated by V14 for different values of $P$ and electron neutrino luminosity $L_{\nu} \equiv (L_{\nu_e} +L_{\bar{\nu}_e})/2$, where $L_{\nu_e}$ and $L_{\bar{\nu}_e}$ are the neutrino and anti-neutrino luminosities, respectively.  As described below, the results for different flux tubes are combined by integrating across the entire open magnetosphere, $\int_{0}^{\theta_{\rm max}}\langle...\rangle d\Omega$, to quantify the total nucleosynthesis of the wind as a function of $P$ and $L_{\nu}$.  The steady-state approximation we adopt is valid if the magnetosphere structure does not change with rotation or due to changes in the convective structure of the star.  The timescale for the outflow to pass through the nucleosynthesis region is much shorter than the timescale over which $L_{\nu}$ is decreasing due to the Kelvin-Helmholtz cooling evolution of the NS, or the timescale over which $P$ is increasing due to angular momentum losses due to magnetic dipole spin-down. 

We start our reaction network calculations just after free nucleons recombine into $\alpha$-particles at a temperature of $T \simeq 7\times 10^{9}$ K.  At small radii, where the magnetic field is dynamically strong compared to the thermal or kinetic energy densities, the force-free approximation is valid and we employ density trajectories from V14.  At sufficiently large radii, matter inertia comes to dominate the energy density of the magnetic field, and the outflows should approach a spherical outflow.  The density profile of a steady-state wind which has reached a constant asymptotic velocity, should approach $\propto 1/r^{2} \propto 1/t^{2}$, while at larger radii where internal velocity gradients become important the profile will approach $\rho \propto 1/r^{3} \propto 1/t^{3}$ appropriate for a freely expanding homologous outflow.  As shown in Figure \ref{fig:approxcube}, we interpolate directly between the V14 trajectories at small radii and the asymptotic $1/t^{3}$ dependence at large radii.  This simplification is motivated by the fact that the qualitative features of the abundance patterns are robust to the detailed density trajectory outside of the radii where charged particle processes cease and the neutron-to-seed ratio has been determined.  

Another simplification is that we neglect heating or deceleration of the wind by the reverse shock, which is produced as the wind interacts with the surrounding supernova ejecta.  This is justified in part\footnote{In addition, highly magnetized winds experience much weaker compressional heating than unmagnetized winds due to the additional support from magnetic pressure.} because the radius of the pulsar wind termination shock $R_{\rm rs} \simeq (v_{\rm ej}/c)^{1/2}R_{\rm ej}$ (e.g.~\citealt{Gaensler&Slane06}), where $v_{\rm ej}$ is the mean velocity of the SN ejecta, is generally much larger than the radius where the formation of seed nuclei occurs.  The latter generally occurs close to the light cylinder radius, which determines the radial scale of outflow divergence.  At time $t$ since explosion we have $R_{\rm ej} \simeq v_{\rm ej}t$ and thus
\be
\frac{R_{\rm rs}}{R_{\rm L}} \approx 37\left(\frac{v_{\rm ej}}{10^{4}\,{\rm km\,s^{-1}}}\right)^{3/2}\left(\frac{t}{1\rm s}\right)\left(\frac{P}{\rm 1\,ms}\right)^{-1}.
\ee
We thus have $R_{\rm rs} \gg R_{\rm L}$ for times $t \gtrsim 1$ s and rotation periods $P \lesssim 10$ ms of interest, implying that seed formation occurs well inside the termination shock.   

\begin{figure}
\begin{center}
\includegraphics[width=1.0\linewidth]{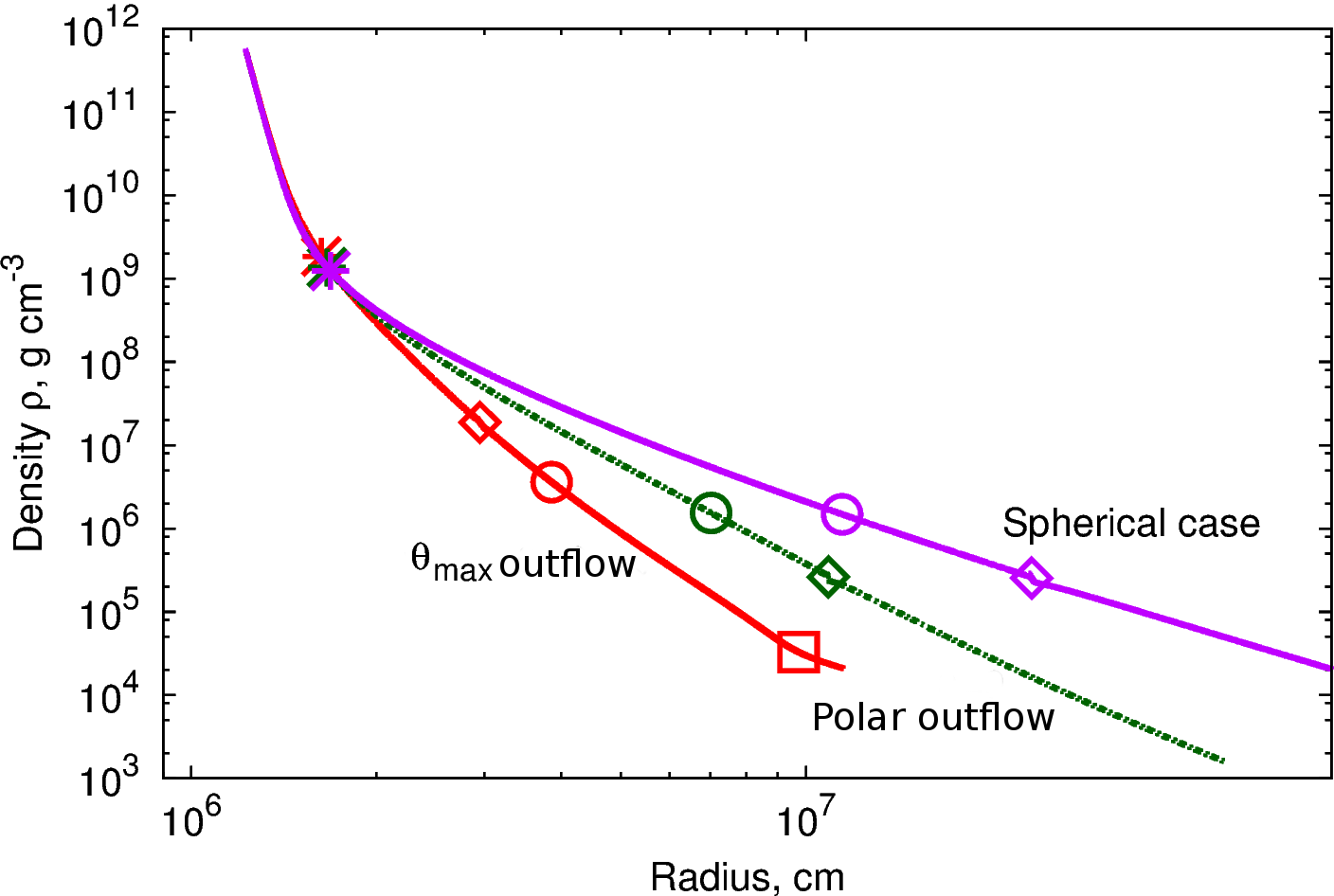}
\caption{Density $\rho(r)$ as a function of the radial coordinate $r$ along the outflow in the neutrino-heated wind of a proto-magnetar with a rotation period $P = 2$ ms and neutrino luminosity $L_{\nu} = 10^{52}$ ergs s$^{-1}$ (V14).  Green and red lines show outflows along polar field lines ($\theta = 0$) and the last open field line ($\theta = \theta_{\rm max}$), while a purple line shows for comparison the trajectory of an otherwise identical spherical wind of the same neutrino luminosity.  Also shown is the point of maximal specific heating (asterisk), approximate location of $\alpha$-particle formation ($T = 0.5$ MeV; circle), the sonic point (diamond), and the light cylinder (square).}
\label{fig:rhoprof}
\end{center}
\end{figure}

\begin{figure}
\includegraphics[width=0.5\textwidth]{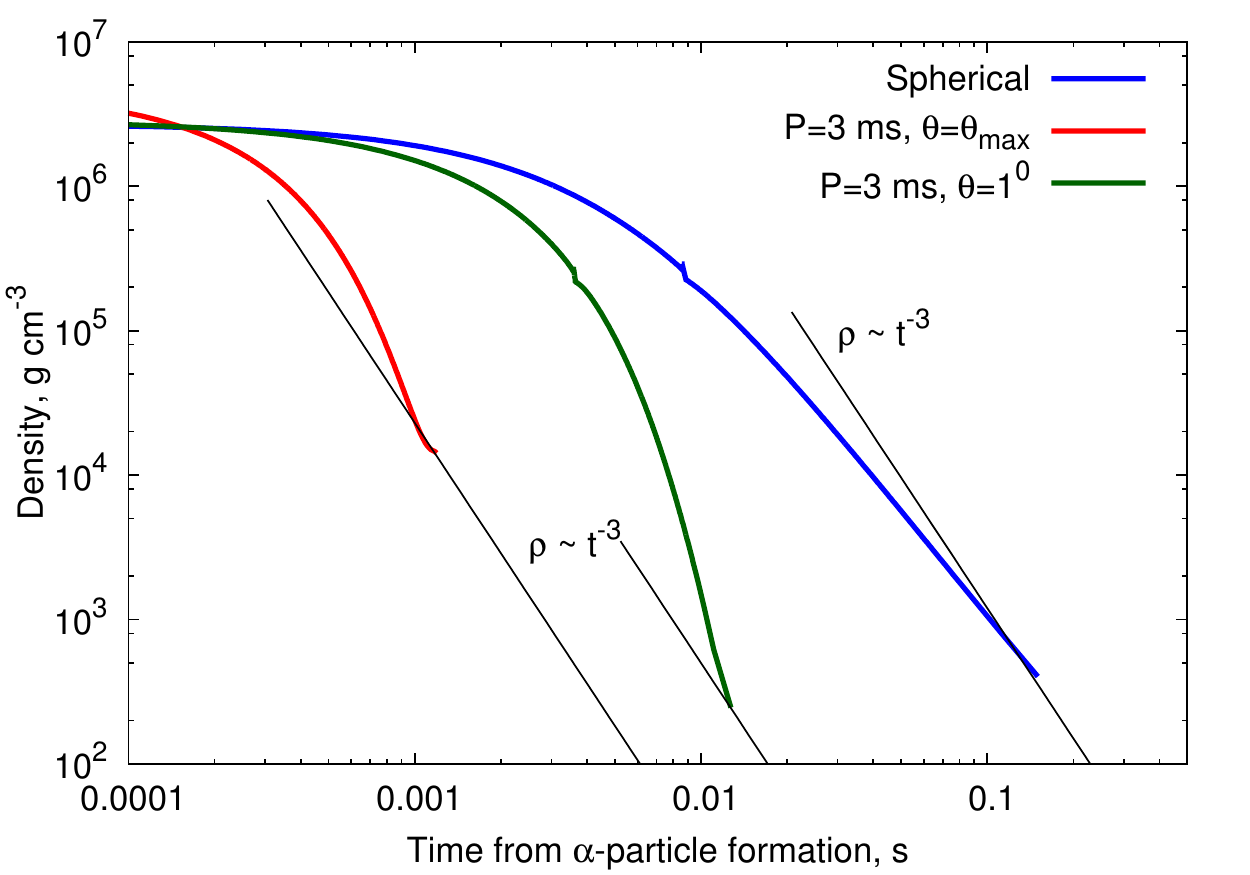}
\caption{Density profiles of wind calculations from V14 for $L_\nu=10^{52}$ ergs s$^{-1}$, for both rotating proto-magnetars with $P=3$ ms (blue and green lines) and spherical wind (blue line).  At large radii we extrapolate the density profile $\rho\propto t^{-3}$ (black lines), as expected at very late times for homologous expansion.}
\label{fig:approxcube}
\end{figure}

We start the nucleosynthesis network after the formation of $\alpha$-particles. By this time, most of the key parameters of the outflow (entropy $s$, expansion timescale $\tau_{\rm dyn}$, electron fraction $Y_e$ and mass loss rate $\dot{M}$) have already been determined in the free nucleon zone at small radii, where nuclear statistical equilibrium (NSE) provides a good approximation for equation of state. Furthermore, after $\alpha$-particle formation, neutrino heating and cooling have become negligible and hence are not included in the network reactions.  Figure \ref{fig:rhoprof} shows that our calculations begin at radii $5-7$ times larger than the location where the neutrino heating rate reaches its maximum just above the NS surface.  The locations of maximum neutrino heating are marked by stars in this figure, while the locations where $T=0.5$ MeV, which approximately corresponds to the locations of $\alpha$-formation, are marked by circles.  

V14 do not account for entropy gain from $\alpha$-particle formation, which we therefore increase by hand from its initial value $s_0$ from V14 according to 
\begin{eqnarray}
 s=s_0+X_\alpha(s,\rho,Y_e)\cdot\De s_\alpha,
\label{eq:s}
\end{eqnarray}
where $X_\alpha(s,\rho)$ is mass fraction of $\alpha$ under NSE with given entropy, $\De s_\alpha=Q_\alpha/T_\alpha$ is entropy gain from $\alpha$ formation, $Q_\alpha$ is the heat of $\alpha$-formation, and $T_\alpha$ is the temperature of $\alpha$ formation.  Since the latter depends weakly on other wind parameters, we adopt a fixed value of $\De s_\alpha=10 k_B$ nucleon$^{-1}$, corresponding to $T_\alpha=8$ GK.  The starting temperature for the reaction network is determined from the density and the entropy according to equation (\ref{eq:s}), with subsequent evolution tracked self-consistently from the density trajectory and radioactive heating.  

Finally, we must specify the initial outflow electron fraction used in our network calculations, which equals the ``final" electron fraction $Y_{e,\rm final}$ set by processes near the PNS surface.  In standard thermally-driven PNS winds, this value is determined by the competition between electron neutrino and electron anti-neutrino capture reactions on free nucleons,
\be
 \nu_e+n \leftrightarrow p+e^- \quad \quad \quad \bar{\nu}_e+p \leftrightarrow n+e^+ \label{eq:captures}.
\ee
These reactions have frozen-out (become slow compared to the expansion rate) at the large radii where our calculations would begin, resulting in an electron fraction close to the equilibrium value set by neutrino absorption reactions (\citealt{Qian&Woosley96})
\be
 Y_{\rm e,eq}=\left(1+\frac{L_{\bar{\nu}_e}}{L_{\nu_e}}\frac{\langle\eps_{\bar{\nu}_e}\rangle-2\Delta+1.2\Delta^2/\langle\eps_{\bar{\nu}_e}\rangle}{\langle\eps_{\nu_e}\rangle+2\Delta+1.2\Delta^2/\langle\eps_{\nu_e}\rangle}\right)^{-1},
\label{eq:Yeeq}
\ee
where $\Delta=1.293$ MeV is proton-neutron mass difference and $\langle\eps_{\bar{\nu}_e}\rangle/\langle\eps_{\nu_e}\rangle$ are the mean energies of electron neutrinos and antineutrinos, respectively.  

For relatively slowly rotating proto-magnetars, $P > 2$ ms, the final electron fraction is very similar to the normal thermally-driven case, i.e. $Y_{\rm e,final} \simeq Y_{\rm e,eq}$.  However, in the most rapidly spinning cases, $P \lesssim 2$ ms, we can have $Y_{\rm e,final} < Y_{\rm e,eq}$ due to rapid magnetocentrifugal acceleration (\citealt{Thompson+04,Metzger+07,Metzger+08b}, V14), which causes a premature freeze-out of the neutrino absorption processes before $Y_e$ is raised completely from its low value near the PNS surface.  More quantitatively, introducing the definition ($\Delta Y_e)_{\rm cent} \equiv Y_{\rm e, final} - Y_{\rm e,eq}$, we find that maximum  ($\Delta Y_e)_{\rm cent} \simeq -0.14$ for $P = 1$ ms, $L_\nu=6\cdot 10^{51}$ ergs s$^{-1}, \theta=\theta_{\rm max}$ and ($\Delta Y_e)_{\rm cent} \simeq -0.011$ for $P = 2$ ms, $L_\nu=10^{52}$ ergs s$^{-1}, \theta=\theta_{\rm max}$. We keep the dependence of $(\Delta Y_e)_{\rm cent}$ on $L_\nu=0.5(L_{\nu_e}+L_{\bar{\nu}_e})$ and $\theta$ and we neglect the dependence on the other parameters ($Y_{e,eq},L_{\nu_e},\eps_{\nu_e},\eps_{\bar{\nu}_e}$).  As the initial conditions for our network calculations we therefore take $Y_{\rm e,final} = Y_{\rm e,eq}(t)+(\Delta Y_e)_{\rm cent}[L_{\nu},\theta]$ for $P=1,2$ ms and $Y_e = Y_e^{\rm eq}(t)$ for all other periods.

The value of $Y_{\rm e,eq}(t)$ is calculated from equation (\ref{eq:Yeeq}) using the time evolution of $L_{\nu_e},L_{\bar{\nu}_e},\langle\eps_{\bar{\nu}_e}\rangle,\langle\eps_{\nu_e}\rangle$ from PNS cooling calculations of \citet{Roberts+12b}. 
Given the uncertainties in neutrino radiation transport (e.g.~\citealt{Fischer+12,Roberts+12b,MartinezPinedo+12}), especially in the essentially unexplored case of very rapid rotation (however, see \citealt{Thompson+05,Thompson07}) and given the effects of ultra-strong magnetic fields on the microphysics (\citealt{Lai&Qian98, Duan&Qian04}), we concentrate on comparing the properties rotating proto-magnetar winds to the conventional unmagnetized spherical wind case.  

\subsection{Reaction Network Calculations}
\label{sec:network}

We start our nucleosynthesis calculations at the point where V14 trajectory reaches $T = 7\times 10^{9}$~K, corresponding to the approximate temperature of $\alpha-$particle formation. However, our actual starting temperature is slightly larger because of the entropy enhancement from $\alpha$-formation (eq.~\ref{eq:s}). We use the nuclear reaction network SkyNet \citep{Lippuner&Roberts15} for the
nucleosynthesis calculation. The composition starts out in nuclear statistical
equilibrium (NSE), a good approximation for the initial temperature $T \gtrsim 7\times 10^9$~K. Given the
extrapolated density as a function of time (see \S\ref{sec:traj}), SkyNet then
evolves the abundances of 7843 nuclear species under the influence of over
140,000 nuclear reactions. The evolved species range from free neutrons and
protons to {${}^{337}$Cn} ($Z = 112$). SkyNet also evolves the entropy and
temperature, which change due to the expansion of the material and energy
released by the nuclear reactions. SkyNet uses a modified version of the Helmholtz
equation of state \citep{Timmes&Swesty00}, which treats every nuclear species as
a separate Boltzmann gas, including also electron-positron gas with arbitrary degree of relativism and photon gas. In some trajectories the initial temperature after taking into account the entropy gain from $\alpha$-formation is above $10^{10}$ K. In these cases, SkyNet uses nuclear statistical equilibrium for evolution of the network up to $10^{10}$ K, below which it switches to full network evolution with all reactions.

The rates of the strong reactions come from the JINA REACLIB database
\citep{cyburt:10}, but only the forward rates are used and the inverse rates
are computed from detailed balance. This is to ensure consistency with NSE,
which depends on the nuclear masses. Spontaneous and neutron-induced fission
rates are taken from \citet{frankel:47}, \citet{panov:10}, \citet{mamdouh:01},
and \citet{wahl:02}. Most of the weak rates come from \cite{fuller:82},
\cite{Oda:94}, and \cite{langanke:00} whenever they are available, and otherwise
the REACLIB weak rates are used. We used the nuclear masses and partition
functions from the WebNucleo XML file distributed with REACLIB,
which contains experimental data where available and finite-range droplet
macroscopic model \citep[FRDM, see e.g.][]{moller:15} data otherwise.

For $Y_e>0.5$, charged particle process nuclei ($Z \simeq 38-40$) can also be formed through $\nu p$ process (e.g.~\citealt{Frohlich+06,Arcones&Montes11}); however, even though SkyNet supports including
neutrino interactions, we do not include these in the present
calculations.  We leave an exploration of charged particle process nuclei formation in proton-rich
proto-magnetar winds to future work.

\subsection{Classes of Abundance Models}
 \label{sec:integ}

\begin{figure}
\begin{center}
\includegraphics[width=1.0\linewidth]{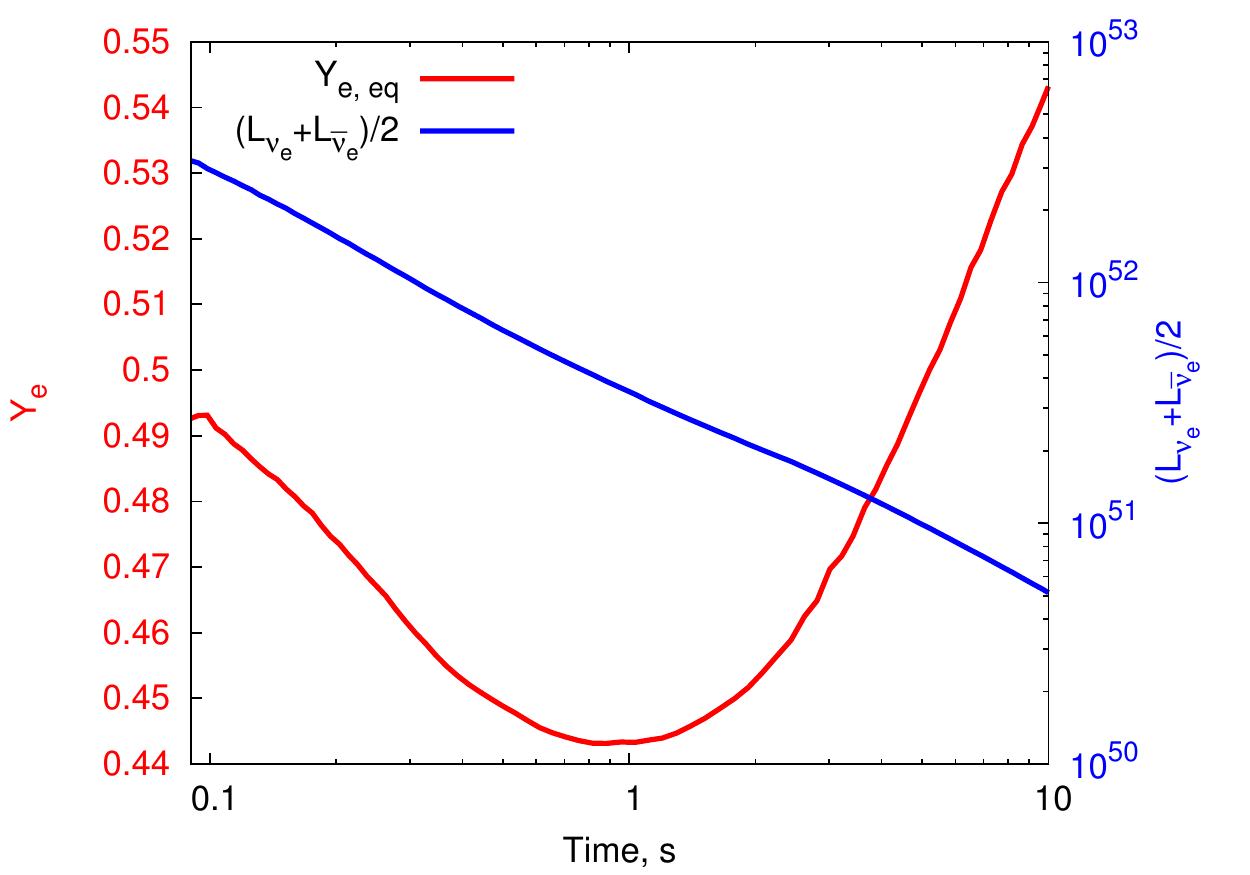}
\caption{The blue line (right axis) shows the time evolution of the electron neutrino luminosity $L_\nu(t) = (L_{\nu_e} + L_{\bar{\nu}_e})/2$ used in our calculations, based on the electron neutrino luminosity $L_{\nu_e}(t)$ and antineutrino luminosity $L_{\bar{\nu}_e}(t)$ from the Kelvin-Helmholtz cooling calculations of \citet{Roberts+12b}.  A red line (left axis) shows the corresponding equilibrium value of the electron fraction $Y_{\rm e,eq}$ (eq.~\ref{eq:Yeeq}) used as input to our nucleosynthesis calculations (we also down-correct $Y_e$ for centrifugal effects for $P \le 2$ ms, but this is not shown here).}
\label{fig:yelnut}
\end{center}
\end{figure}

The nucleosynthesis products of proto-magnetar winds vary as a function of the latitude of the open flux tube at a fixed time.  They also vary in a global sense, integrated over the entire magnetosphere at a fixed time, or averaged over all times in the PNS cooling evolution.  We cover the range of possible diagnostics of the wind nucleosynthesis by calculating the abundance patterns in three general cases:
\begin{enumerate}
\item Individual flux tubes along different latitudes $\theta$ at a fixed time (or equivalently, neutrino luminosity), producing abundance yields as a function of $(\theta,L_\nu,Y_e,P$). \\
\item The entire wind at a fixed time, by integrating individual flux tubes over the solid angle of the open magnetosphere, producing abundance yields as a function of ($L_\nu,Y_e,P$).  \\
\item The entire wind (integrated over the open magnetosphere) averaged over the Kelvin-Helmholtz cooling epoch $L_{\nu}(t)$ (Fig.~\ref{fig:yelnut}), producing abundance yields as a function of just the rotation period $P$.    \\
\end{enumerate}
In the angle-integrated cases, we weight the abundances by the mass loss rate along each given open flux tube.  In the time-integrated case, they are weighted also by the total ejected mass from the entire magnetosphere at each epoch $t$.  In the latter case, by fixing the rotation period, we have assumed that the magnetic spin-down time of the pulsar is longer than Kelvin-Helmholtz cooling timescale of seconds; this approximately is justified for surface magnetic dipole field strengths of $B_{\rm d} \lesssim 10^{16}$ G. 

In the time-integrated case, we also consider two methods for treating the time evolution of the electron fraction $Y_e$.  In one case, we fix $Y_e$ at a constant value throughout the cooling epoch.  In the second case we derive its value as described in $\S\ref{sec:traj}$ using the equilibrium electron fraction (eq.~\ref{eq:Yeeq}) from the time evolution of $L_\nu,L_{\bar{\nu}},\langle\eps_{\bar{\nu}_e}\rangle,\langle\eps_{\nu_e}\rangle$ from the PNS cooling calculations of \citet{Roberts+12b}, as shown on Figure \ref{fig:yelnut}.  Since we are focused on the $r$-process, we integrate only over epochs when $Y_{e,eq}<0.5$, i.e.~at times $0.1\lesssim t\lesssim 5$ s, and all reported wind properties (e.g., ejecta mass, mass fractions, etc. - all the data in Table \ref{table:summary}) refer to just this time period.  This is a reasonable approximation to the total yield of the wind, since the bulk of the total mass loss is occurs at early times $t \sim 1-2$ s. 
\section{Results \label{sec:results}}
\label{sec:results}
\subsection{Variation in wind properties across the magnetosphere}

Based on the $s^3/\tau_{\rm exp}$ threshold criterion of \citet{Hoffman+97}, V14 concluded that rotating proto-magnetars are more suitable for producing heavy $r$-process nuclei than in the spherical wind case, but not to the degree necessary to reach the third abundance peak ($Z \gtrsim 78$) for currently favored values of $Y_e \gtrsim 0.4$.\footnote{For some trajectories in the most rapidly spinning $P=1$ ms case for which $Y_e \lesssim 0.38$, the value of $s^3/(\tau_{\rm exp} Y_e^3)$ did exceed the modified \citet{Hoffman+97} criterion for reaching $A = 200$.}  Although our nucleosynthesis calculations presented here confirm this broad conclusion, we find some quantitative differences.  While V14 found that the most promising rotation period for producing heavy elements was $P = 2$ ms, here we find that $P = 1$ ms is instead optimal. $P=1$ ms is a special case because of centrifugal effects are particularly pronounced, and the values of $s$ and $Y_{\rm e,final}$ thus differ significantly from the spherical case.   

We focus our analysis in this section on models with $P = 1$ and $3$ ms.  The latter case plays important role because the rotational energy of the magnetar with $P = 3$ ms, $E_{\rm rot} = I \Omega^{2}/2 \approx 10^{51}$ ergs, is comparable to the kinetic energy of a normal SN.  Hence magnetars born with $P \gtrsim 3$ ms could in principle be `hidden' among the normal population of normal SNe without violating constraints on the observed kinetic energies of their ejecta from SN spectra and the total energies of their remnants (as already discussed, hypernovae are known to be rare; e.g.~\citealt{Podsiadlowski+04,Woosley&Bloom06}).

\begin{figure}
\subfigure{
\includegraphics[width=0.5\textwidth]{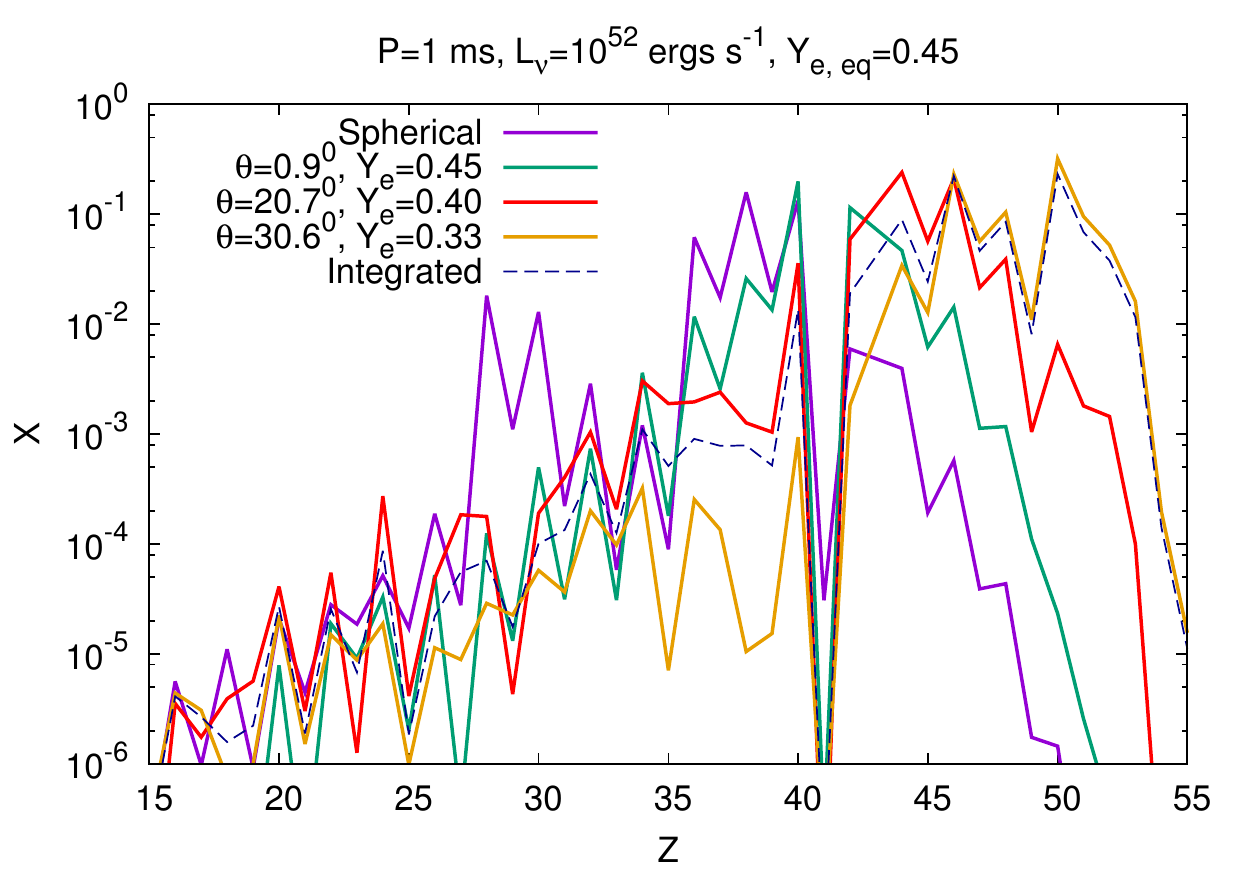}}
\subfigure{
\includegraphics[width=0.5\textwidth]{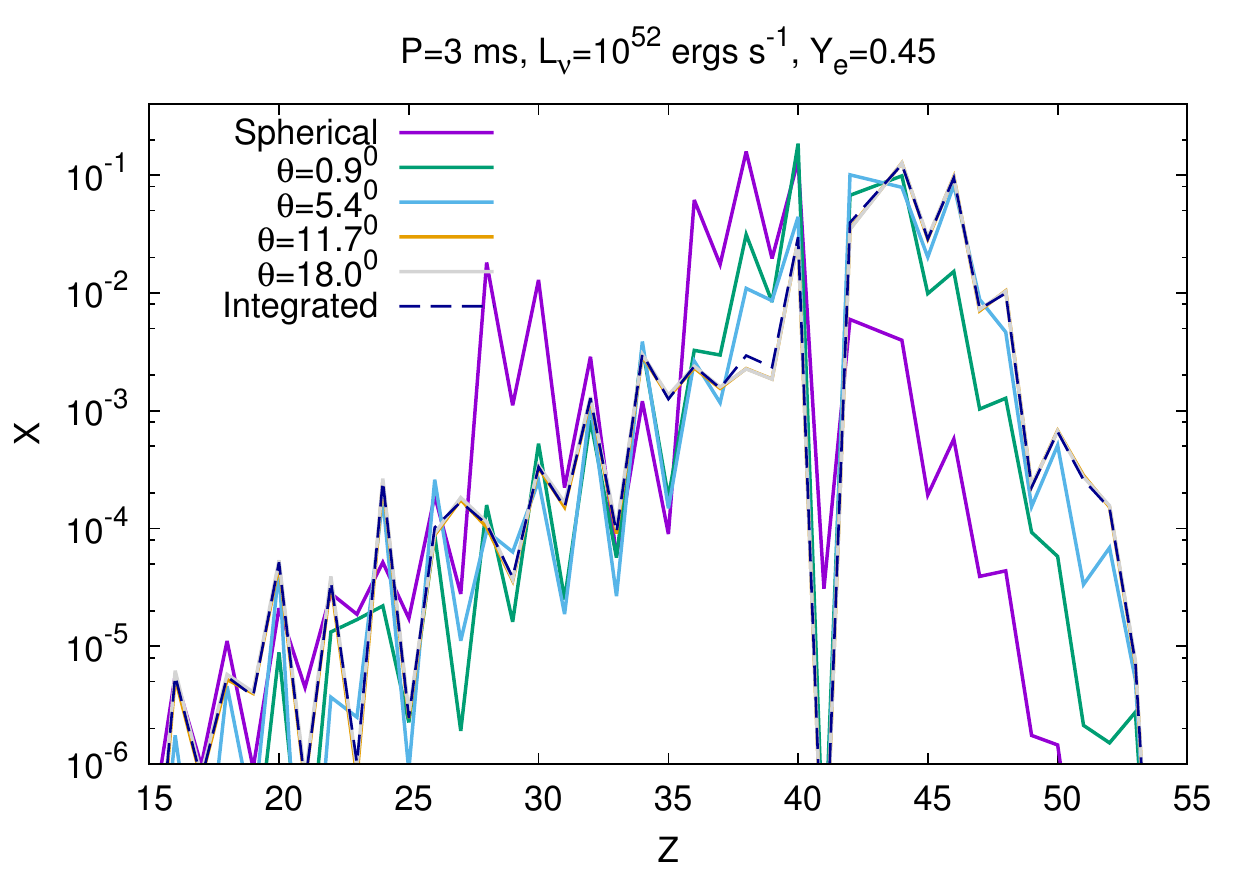}}
\caption{Mass fraction of nuclei $X(Z)$ synthesized in the proto-magnetar wind of neutrino luminosity $L_\nu=10^{52}$ ergs s$^{-1}$, and neutrino equilibrium electron fraction $Y_e = 0.45$.  Results are shown for two rotation periods: $P = 1$ ms (top panel) and $P = 3$ ms (bottom panel); in the former case, the value of $Y_e$ is lower than its equilibrium value set by neutrino absorptions due to centrifugal acceleration effects, as marked in the key.  Different lines correspond to the abundances synthesized in the wind along field lines with different polar angles $\theta$ at the NS surface, ranging from near the pole ($\theta \simeq 0$) to the last open field line $\theta_{\rm max}$ (depending on the rotation period). Shown for comparison with a blue dashed line are the mass-weighted abundances integrated over the solid angle of the open magnetosphere.  A purple line shows the abundances from an otherwise identical calculation of a spherical non-rotating non-magnetized wind.}
\label{fig:thetanuc}
\end{figure}

Figure \ref{fig:thetanuc} shows the mass fraction as a function of nuclear charge $Z$ for different outflow latitudes $\theta$, calculated for $P=1$ ms (top panel) and $P = 3$ ms (lower panel).  For all values of $\theta$, the abundance distribution extends to higher masses $Z$ than the otherwise equivalent spherical wind with the same $L_{\nu}$ and $Y_e$, which is shown for comparison with a purple line.  Part of this effect is due to the shorter expansion time through the seed formation region caused by the faster diverging outflow areal function $\propto 1/r^{3}$ in the dipole magnetic field, as compared to the spherical wind case (areal function $\propto 1/r^{2}$), as well as centrifugal force from rotation.  

Figure \ref{fig:thetanuc} also shows that the abundance distribution proceeds to heavier elements with increasing  $\theta$, due to the additional acceleration caused by magneto-centrifugal acceleration along field lines inclined with respect to the rotation axis. The heaviest nuclei are synthesized in those outflows which graze the closed zone and pass near the equatorial plane outside the light cylinder. 

Shown for comparison with a dashed blue line are the abundances integrated over the open zone of the entire magnetosphere, from which it is apparent that the flow properties near the last open field line ($\theta_{\rm max}$) also dominate the total abundance of the wind.  This is expected because outflows with larger $\theta$ contribute a greater fraction of the total open solid angle of the magnetosphere and, to a lesser extent, because the mass loss rate per unit surface area is enhanced by magneto-centrifugal acceleration for larger $\theta$ (\citealt{Thompson+04,Metzger+07}).  

\begin{figure}
\includegraphics[width=0.5\textwidth]{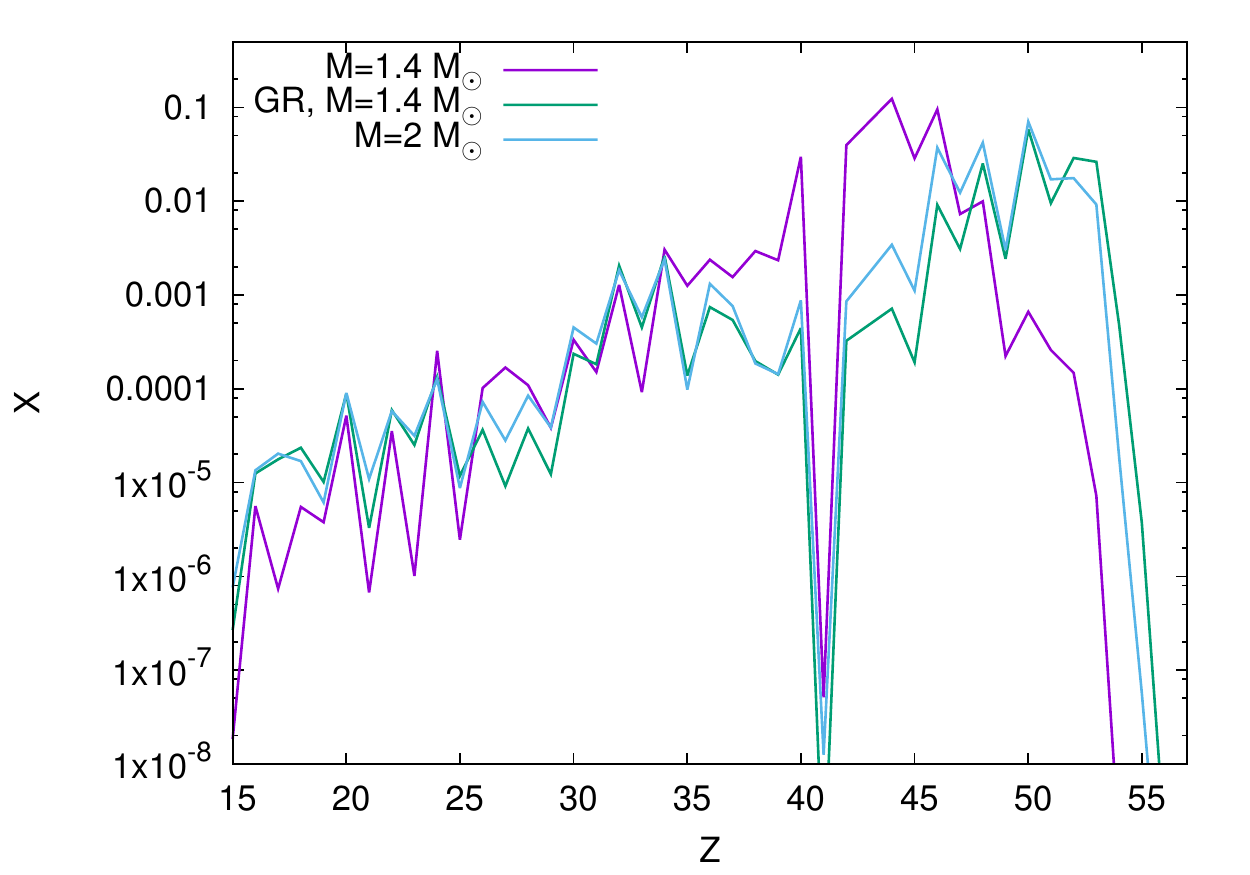}
\caption{Mass fraction of nuclei, integrated across the entire open magnetosphere of a proto-magnetar with $P = 3$ ms, $Y_e=0.45$, and $L_{\nu} = 10^{52}$ ergs s$^{-1}$.  Here we compare the fiducial case of Newtonian gravity for a 1.4$M_{\odot}$ NS (purple line) to the result for a $2.0M_{\odot}$ NS (blue line).  Also shown is the case of a 1.4$M_{\odot}$ NS (green line) calculated using a modified Paczy\'{n}ski-Wiita potential, meant to mimic the deeper potential well of GR.  } 
\label{fig:compnuc}
\end{figure}

Most of our calculations employ Newtonian gravity for a NS of mass $M=1.4 M_\odot$.  However, Figure \ref{fig:compnuc} shows the results for a more massive NS with $M=2 M_\odot$ NS, as well as for a model with $M=1.4 M_\odot$ but using the Paczy\'{n}ski-Wiita potential to mimic the effects of general relativity (GR).  The effect of a higher NS mass, or the {\it effectively} higher mass due to the deeper Paczy\'{n}ski-Wiita potential, is also to increase the maximum mass nuclei synthesized, extending it up to $Z \approx 54$ (Xenon), near the peak of the 2nd $r$-process peak.  This well known effect results because of the additional heating, and hence higher asymptotic entropy, achieved by the winds to escape from the deeper potential well (\citealt{Cardall&Fuller97}).  Note, however, that we have not included the GR-induced gravitational redshift on the mean neutrino energies, which somewhat reduces the neutrino heating rate and acts to mitigate this effect (e.g.~\citealt{Thompson+01}).

\subsection{Time-integrated models}
\begin{table}
\begin{scriptsize}
\begin{center}
\vspace{0.05 in}\caption{Summary of Time-integrated Models}
\label{table:summary}
\begin{tabular}{ccccccc}
\hline 
\hline
P, ms & $\bar{A}^{(a)}$ & $\bar{Z}^{(a)}$ & $X_\alpha$ & $X_n^{\dagger}$ & $X_p^{\dagger}$ & $M_{\rm ej},\quad M_{\odot}$ \\
-$^{*}$ & 87.34 & 37.71 & 0.561 & $ 9.24 \cdot 10^{-7} $  & $ 3.65 \cdot 10^{-7} $  & $ 9.98 \cdot 10^{-4} $  \\
1 & 98.94 & 42.61 & 0.429 & $ 4.83 \cdot 10^{-4} $  & $ 5.58 \cdot 10^{-5} $  & $ 3.45 \cdot 10^{-3} $  \\
2 & 98.73 & 42.53 & 0.685 & $ 3.42 \cdot 10^{-4} $  & $ 4.07 \cdot 10^{-5} $  & $ 1.79 \cdot 10^{-4} $  \\
3 & 97.81 & 42.13 & 0.719 & $ 2.18 \cdot 10^{-4} $  & $ 2.55 \cdot 10^{-5} $  & $ 1.06 \cdot 10^{-4} $  \\
4 & 97.36 & 41.94 & 0.714 & $ 1.90 \cdot 10^{-4} $  & $ 2.22 \cdot 10^{-5} $  & $ 7.80 \cdot 10^{-5} $  \\
5 & 95.07 & 40.96 & 0.688 & $ 1.62 \cdot 10^{-4} $  & $ 1.90 \cdot 10^{-5} $  & $ 6.64 \cdot 10^{-5} $  \\
10 & 94.07 & 40.53 & 0.675 & $ 9.34 \cdot 10^{-5} $  & $ 1.08 \cdot 10^{-5} $  & $ 2.52 \cdot 10^{-5} $  \\

\hline
\hline
\end{tabular}
\end{center}
$^{(a)}$ Mean mass number for all elements except H and He; $^{*}$Spherical solutions; $^{\dagger}$ Mass fraction of free neutrons and free protons at $t=100$ s, i.e. prior to the decay of free neutrons.  
\end{scriptsize}
\end{table}

\begin{figure}
\subfigure{
\includegraphics[width=0.5\textwidth]{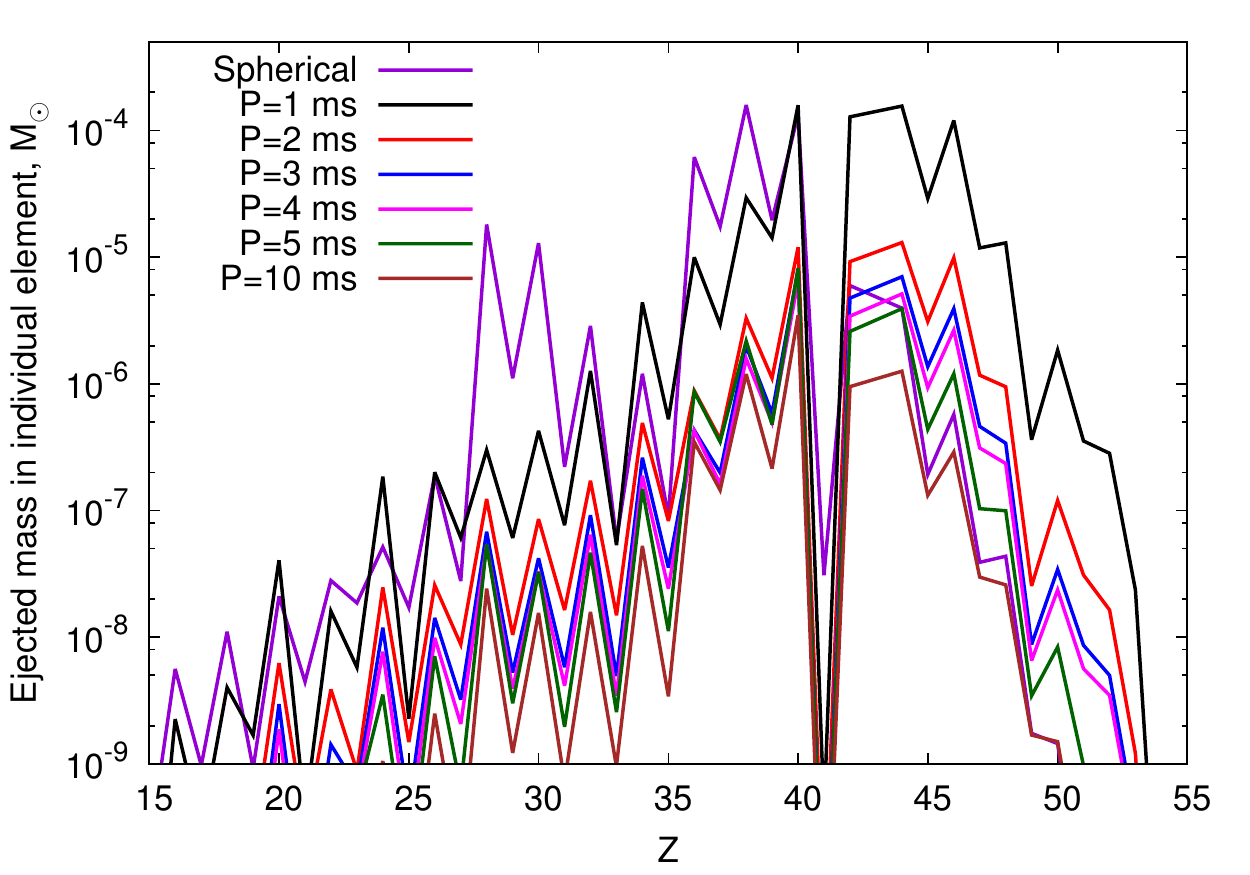}}
\subfigure{
\includegraphics[width=0.5\textwidth]{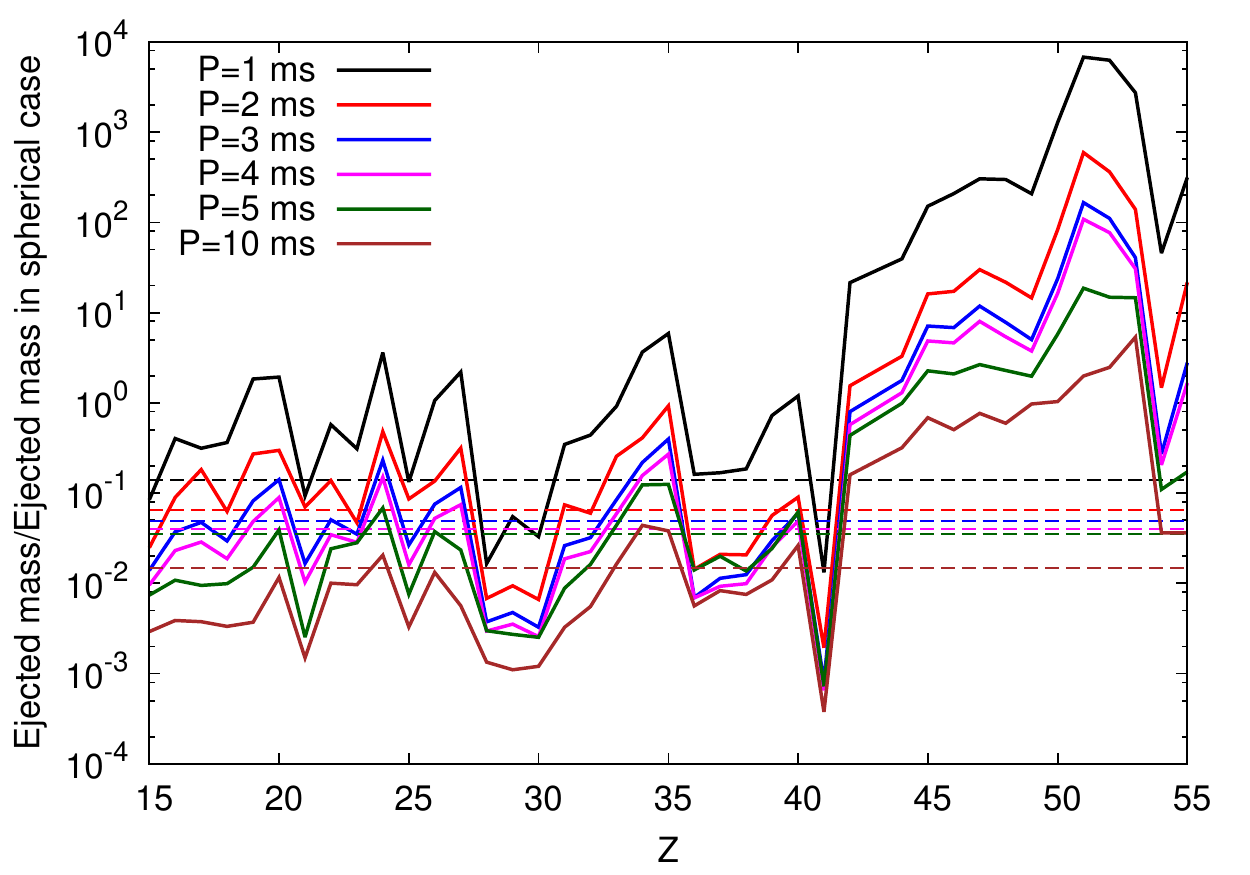}}
\caption{{\it Top}: Total wind ejecta mass in individual elements, time-integrated across the Kelvin-Helmholtz cooling epoch using the $Y_{e}(t)$ evolution from \citet{Roberts+12b} (Fig.~\ref{fig:yelnut}) and $(\De Y_e)_{\rm cent}$ from V14 for $P=1,2$ ms. {\it Bottom:} Ratio of the wind ejecta mass of each element in proto-magnetar models relative to those in the otherwise equivalent spherical non-rotating wind models. The dashed lines of same color shows the purely geometric factor $f_{\rm open}$ (eq.~\ref{eq:fopen}) arising from fraction of the PNS magnetosphere open to outflows.} 
\label{fig:elmej}
\end{figure}

Figure \ref{fig:elmej} shows the total yield of proto-magnetar winds, integrated across the entire Kelvin-Helmholtz cooling evolution using the $Y_{e}(t)$ evolution from \citet{Roberts+12b} (Fig.~\ref{fig:yelnut}) and $(\De Y_e)_{\rm cent}$ from V14 for $P=1,2$ ms. We show both the total mass (top panel) as well as in ratio of abundances to those calculated in the otherwise equivalent case of a spherical wind (bottom panel). One clear trend is that the abundance distribution extends to heavier elements in the rotating case, with a larger number of heavy elements synthesized with decreasing rotation period as compared to the otherwise equivalent spherical case.  

However, strong magnetic fields also {\it reduce} the mass of the ejecta in light elements.  This reduction is due to the fact that only a small fraction of the PNS surface is open to outflows, such that the total ejecta mass from each event is typically smaller than the spherical case by this purely geometric factor of 
\be f_{\rm open} \approx \frac{2\pi \theta_{\rm max}^{2}}{4\pi} \approx \frac{1}{2}\frac{R_{\rm ns}}{R_{\rm Y}} \approx 0.13\left(\frac{P}{1\,\rm ms}\right)^{-1},
\label{eq:fopen}
\ee
which we show as a dashed line in Fig.~\ref{fig:elmej}.  The only period that exhibits a significant enhancement compared to this geometric factor across most elements is $P = 1$ ms because of the large centrifugal enhancement in the mass loss rate per unit surface area, which overcomes the (comparatively modest) reduction due to $f_{\rm open} \approx 0.1$ in this fastest spinning case. 

\begin{figure}
\includegraphics[width=0.5\textwidth]{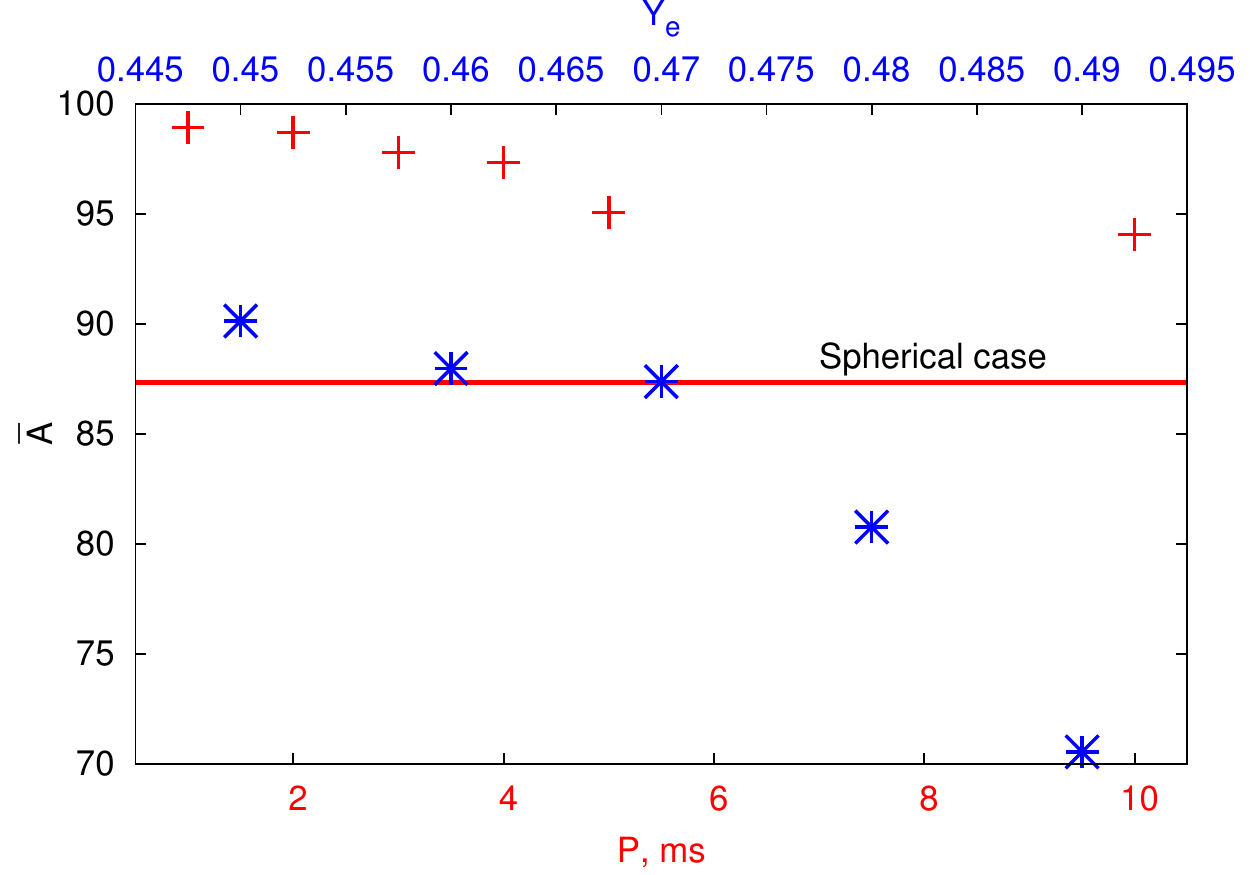}
\caption{Abundance-weighted mass $\bar{A}$ of synthesized nuclei, excluding H and He. {\it Bottom Axis}: Red crosses show the results for time-integrated models with $Y_e(t)$ evolution from \citet{Roberts+12b} (Fig.\ref{fig:yelnut}) for different rotation periods, while the solid line shows the result for the non-rotating spherical wind case for the same $Y_e$ evolution. {\it Top Axis:}  Blue asterisks show time-integrated models of a spherical wind which instead assume a temporally constant value of $Y_e$ as marked on the top axis.}
\label{fig:abar}
\end{figure}

Figure \ref{fig:abar} shows the abundance-weighted mass $\bar{A}$ of synthesized nuclei as a function of the magnetar rotation period (red crosses, bottom axis), from which it is clear that $\bar{A}$ decreases monotonically with increasing spin period $P$.  For comparison a red line shows the result for a spherical wind using the same neutrino-cooling evolution from \citet{Roberts+12b} as in the magnetized case. The mean value of $\bar{A}$ for $P=1-10$ ms is 97, but the $\bar{A}$ in spherical case is $\bar{A}_{\rm sph}=87$: hence, the rotation with $P=1-10$ ms lowers $\bar{A}$ by 10 units. On the top axis we also show the value of $\bar{A}$ calculated in the spherical non-rotating case assuming a temporally {\it fixed} value of $Y_e$. 

 The fact that the ejecta mass in individual elements exceeds those produced in the spherical case by a factor up to $10^2$ (or $10^3$ in case of $P=1$ ms) has the striking implication that millisecond magnetars possess unique nucleosynthetic signatures, which would be measurable even if only one in $100(1000)$ neutron stars were born strongly magnetized with $P\lesssim 4$ ms ($P\approx1$ ms).

Table \ref{table:summary} summarizes important quantities from our time-integrated models.  We see that centrifugal effects enhance the values of $\bar{A}$ and $\bar{Z}$ moving to shorter periods.  Enhanced centrifugal mass loss overpowers the geometric factor for $P=1$ ms, such that the total ejecta mass of $M_{\rm ej} \approx 3\times 10^{-3}M_{\odot}$ for $P=1$ ms is about three times higher than in the spherical case. 

\section{Discussion}
\label{sec:discussion}

\subsection{Rate constraints on the birth of magnetars}

If millisecond proto-magnetars are found to produce large quantities of rare isotopes, one could in principle place constraints on their birth periods based on their nucleosynthetic yields compared to those in our Galaxy as inferred from the solar abundances.  An upper limit on birth rate of magnetars of a given rotation period is given by
\be
 R\le {\rm min}\frac{X_{{\rm i},\odot}M_{\rm gas}}{M_{\rm ej, i}t}=\frac{X_{{\rm i},\odot}M_{\rm gas}}{X_{\rm i, wind}M_{\rm ej}t_{\rm gal}} \label{eq:rates},
\ee
where $X_{\rm i, wind}$ is mass fraction of an element (or an isotope) in the PNS wind, $M_{\rm ej}$ is the total wind ejecta mass, $X_{{\rm i},\odot}$ is the mass fraction of the same element (or isotope) on the Sun, $M_{\rm gas}\approx 3.3 \times 10^{10} M_\odot$ is the total mass of gas in the Galaxy when the Sun formed, and $t_{\rm gal} \approx 10^{10}$ years is the age of the Galaxy when the Sun formed.  In other words, the total ejecta mass of a given element or isotope under consideration cannot exceed the total mass of that element contained in the gas from which the Sun formed (assuming the Sun formed from ``ordinary'' gas with abundances representative of the Galactic mean).

\begin{figure}
\includegraphics[width=0.5\textwidth]{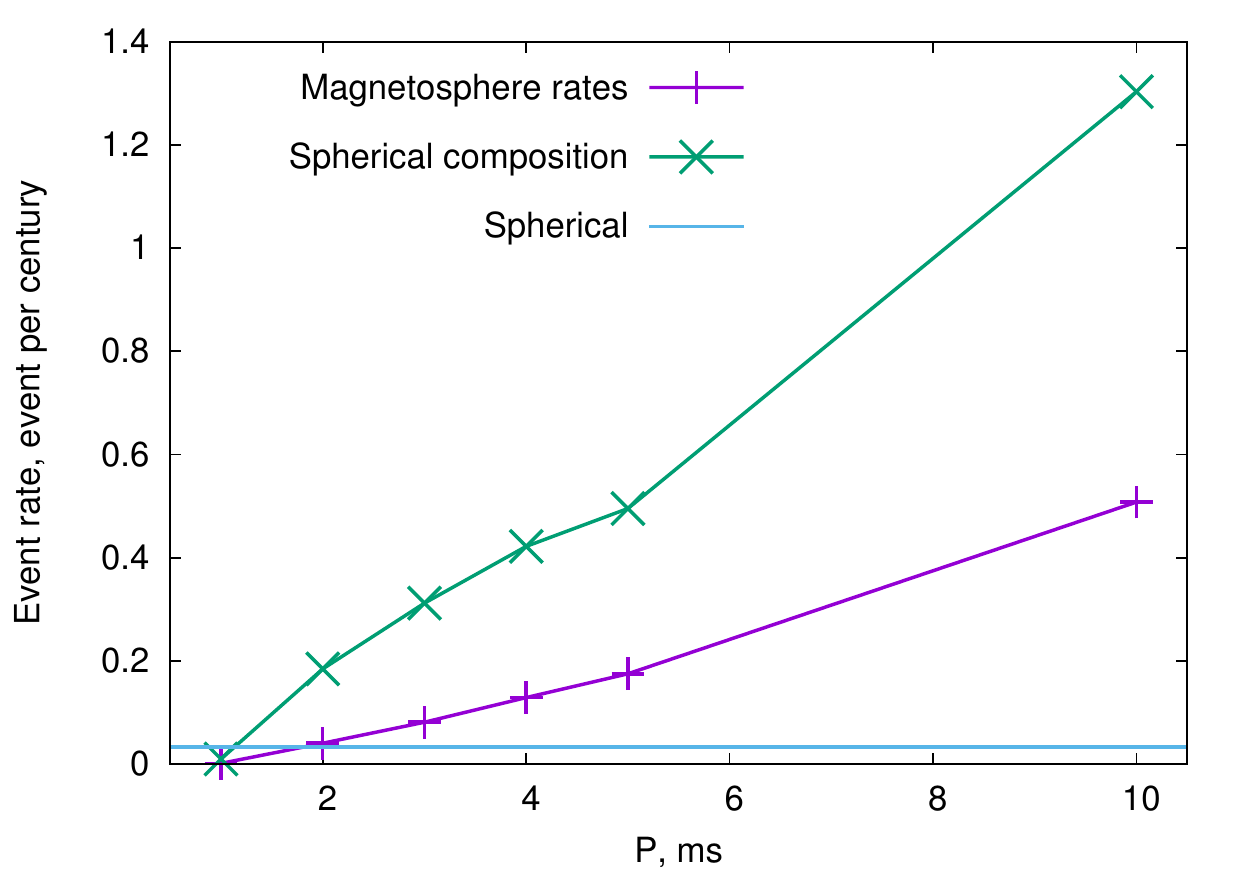}
\caption{Upper limit on rates of magnetar birth as a function of the birth rotation period $P$ so as not to overproduce solar system abundances of $r$-process nuclei (eq.~\ref{eq:rates}).  Purple crosses show the results based on our abundance calculations of proto-magnetar winds (Fig.~\ref{fig:elmej}).  Green crosses show the rate constraint that would result if we used the composition from the spherical model, but down-correcting the ejecta mass $M_{\rm ej}$ to account for the purely geometric correction $f_{\rm open}$ (eq.~\ref{eq:fopen}) resulting from the small solid angle of the open magnetosphere.  
} 
\label{fig:rates}
\end{figure}

Figure \ref{fig:rates} shows the result of such a calculation of the maximum allowed birth rate in events per century, which is approximately also the fraction of neutron star births.  A purple line shows the rate derived from our time-integrated calculation of proto-magnetar winds, as a function of the magnetar rotation period.  Shown for comparison with a blue line is the rate limit for the standard non-rotating spherical wind case.  

It may be surprising that, for standard spherical winds, the allowed event rate is very low,  $\lesssim 0.05$ per century.  This is due to the well-known fact that standard spherical winds overproduce the charged particle process nuclei with $Z = 38-40$ (e.g.~\citealt{Arcones&Montes11}) for the neutron-rich wind conditions $Y_e \lesssim 0.5$ found using contemporary PNS cooling calculations (\citealt{Roberts+12b,MartinezPinedo+12,Fischer+12}). 

It may also be surprising that the rate constraints are weaker on proto-magnetar winds than on spherical winds, despite the fact that the nucleosynthesis of proto-magnetars extends to higher $Z$ elements, which are rarer in the solar system.  This is again explained by the fact that only a small fraction of the PNS surface is open to outflows, such that the total ejecta mass from each event is typically smaller than the spherical case by the geometric factor $f_{\rm open}$ (eq.~\ref{eq:fopen}).  The green line in Fig.~\ref{fig:rates} shows the rate constraints one would derive if the composition of a spherical wind was attenuated by the purely geometric factor $f_{\rm open}$.  The resulting rate constraint is weaker than the proto-magnetar case including the full effects of the magnetized wind dynamics on the composition itself (purple line); this shows that the rare higher $Z$ elements which are synthesized proto-magnetar winds do tighten the rate constraint significantly, but not enough to overcome the purely geometric suppression factor $f_{\rm open}$.  

The net result of all of this is that the effects of magnetic fields and rotation are sufficiently modest that - given also current uncertainties in the electron fraction of the wind - one cannot at present place meaningful constraints even on the birth rate of non-rotating PNS, much less on their birth periods and magnetic field strengths.  Still, our results show that at least under our assumption for the electron fraction of the wind, the birth rate of magnetars {\it with millisecond periods} cannot exceed $\sim 1-10\%$ of the core collapse SNe rate, depending on rotation period.  Reassuringly, this number exceeds the {\it total} birth rate of Galactic magnetars (\citealt{Woods&Thompson06}) and is consistent with the lower millisecond magnetar birth rate inferred if they power hydrogen-poor superluminous supernovae (e.g.~\citealt{Quimby+13}).

It is possible that neutron stars are generically born with magnetar-strength fields.  This could occur, for instance, if strong fields are generated by a convective dynamo in the PNS, which later decay away in the majority of cases by the time they are observed as radio pulsars.  We note that, in such a case the reduction in the open fraction of the PNS surface due to confinement by a dynamically-important magnetic fields (\citealt{Thompson03}) would be generic to all winds and thus could help alleviate the current overproduction of the charged particle process nuclei $Y_e < 0.5$ wind models (even if rapid rotation itself is comparatively rare).  Also note that including GR will increase the depth of potential well for matter to escape and lower neutrino energy because of gravitational redshift.  Both these factors will lower the ejecta mass, also alleviating constraints on the event rate.

\subsection{As sources of light r-process nuclei in metal-poor stars}
\label{sec:weak}

\begin{figure}
\includegraphics[width=0.5\textwidth]{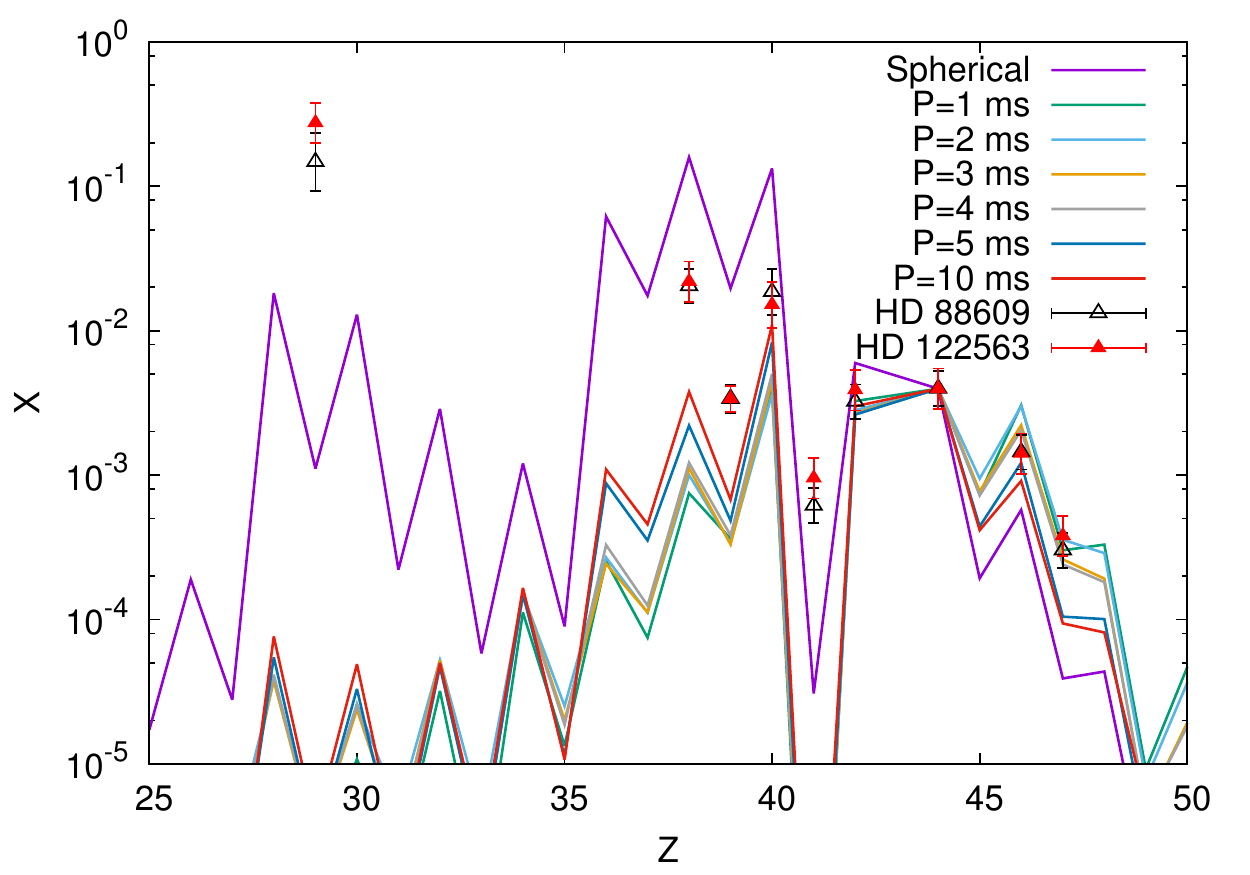}
\caption{Abundances in time-integrated models compared to abundances of Galacic metal-poor stars HD 122563 (\citealt{Honda+06}) and HD 88609 (\citealt{Honda+07}).  Abundances are normalized to the Ru ($Z=44$) abundance of our spherical time-integrated model. After $Z=55$, our models produce very small abundances and they are not able to match the abundances observed in HD 122563 and HD 88609.}
\label{fig:hd}
\end{figure}

Figure~\ref{fig:hd} shows the abundances of our time-integrated models (Fig.~\ref{fig:elmej}) compared to the abundances of two Galactic metal-poor stars, HD 122563 (\citealt{Honda+06}) and HD 88609 (\citealt{Honda+07}) which show a relative dearth of ``heavy" $r$-process nuclei ($Z \gtrsim 56$) compared to the solar system; all abundances have been normalized to the Ru ($Z=44$) abundance in our spherical model.  Although the spherical wind models produce large quantities of charged particle process nuclei ($Z = 38-40$), they underproduce the solar abundances of the weak r-process nuclei ($Z = 41-55$) in HD 122563 and HD 88609.  By contrast, our rotating magnetar models produce larger abundances of weak $r$-process nuclei with $Z \gtrsim 40$, while comparatively underproducing the lighter charged particle process nuclei. 

Our results provide a potential explanation, within the PNS wind paradigm, for the fact that  light $r$-process nuclei with $Z \lesssim 56$ in metal-poor stars show greater star-to-star variation than the heavier $r$-process nuclei (e.g.~\citealt{Roederer+10}).  We propose that such star-to-star variation could be understood if the charged particle process and weak $r$-process nuclei are produced by a combination of normal (slowly-rotating and/or weakly magnetized) NSs and a subclass of magnetars with a range of birth rotation periods.  

Magnetars with very short birth rotation periods of $P \lesssim 1$ ms have also been discussed as a source of heavy $r$-process nuclei, mainly by the ejection of low-$Y_e$ matter during the early dynamical phases of the bipolar explosion (\citealt{Metzger+08b,Winteler+12,Nishimura+15}).  Although such extreme magnetars are probably rare, potentially disfavoring this channel as the dominant Galactic source of heavy $r$-process nuclei, V14 point out that magnetars with less extreme birth periods of $\sim 2-5$ milliseconds could produce light $r$-process nuclei in a larger fraction of events.  The rapid core rotation rates of the massive progenitor stars giving rise to millisecond proto-magnetars are also likely to be more common at the low metallicities which characterized early epochs in the chemical evolution of our Galaxy (\citealt{Stanek+06}).

Both this work and V14 focused on the winds from aligned proto-magnetars, for which the rotation axis coincides with magnetic axis.  Studies of the more general case of inclined rotators are less developed because the problem is inherently non-stationary (for review, see e.g.~\citealt{Cerutti&Beloborodov16}).  Numerical studies have shown that the equatorial region of the magnetar outflow in this case is characterized by `stripes' of alternating magnetic fields, which are separated by current sheets and are prone to magnetic reconnection (e.g.~\citealt{Spitkovsky06}).  Dissipation of magnetic energy, for instance as these stripes reconnect outside the light cylinder radius (e.g.~\citealt{Lyubarsky&Kirk01}), provides an additional possible source of heating in the wind, which would act to both increase the entropy of the flow and contribute to its acceleration.  If this heating occurs near the seed formation radius, typically close to the light cylinder in our models, this would substantially enhance the prospects for synthesizing third-peak $r$-process nuclei in misaligned rotators compared to the aligned case studied here.  The magnetic inclination angle therefore provides an another parameter, in addition to the rotation period, which could impart diversity to the $r$-process yields of proto-magnetar winds.  

More work is clearly needed to distinguish the magnetar hypothesis from other proposed sites of the light $r$-process elements, such as binary neutron star mergers (\citealt{Fernandez&Metzger13,Wanajo+14,Perego+14,Just+15,Goriely+15,Martin+15b,Wu+16}).

\subsection{Gamma-ray burst engines and UHECR sources} \label{sec:uhecr}

\begin{figure}
\includegraphics[width=0.5\textwidth]{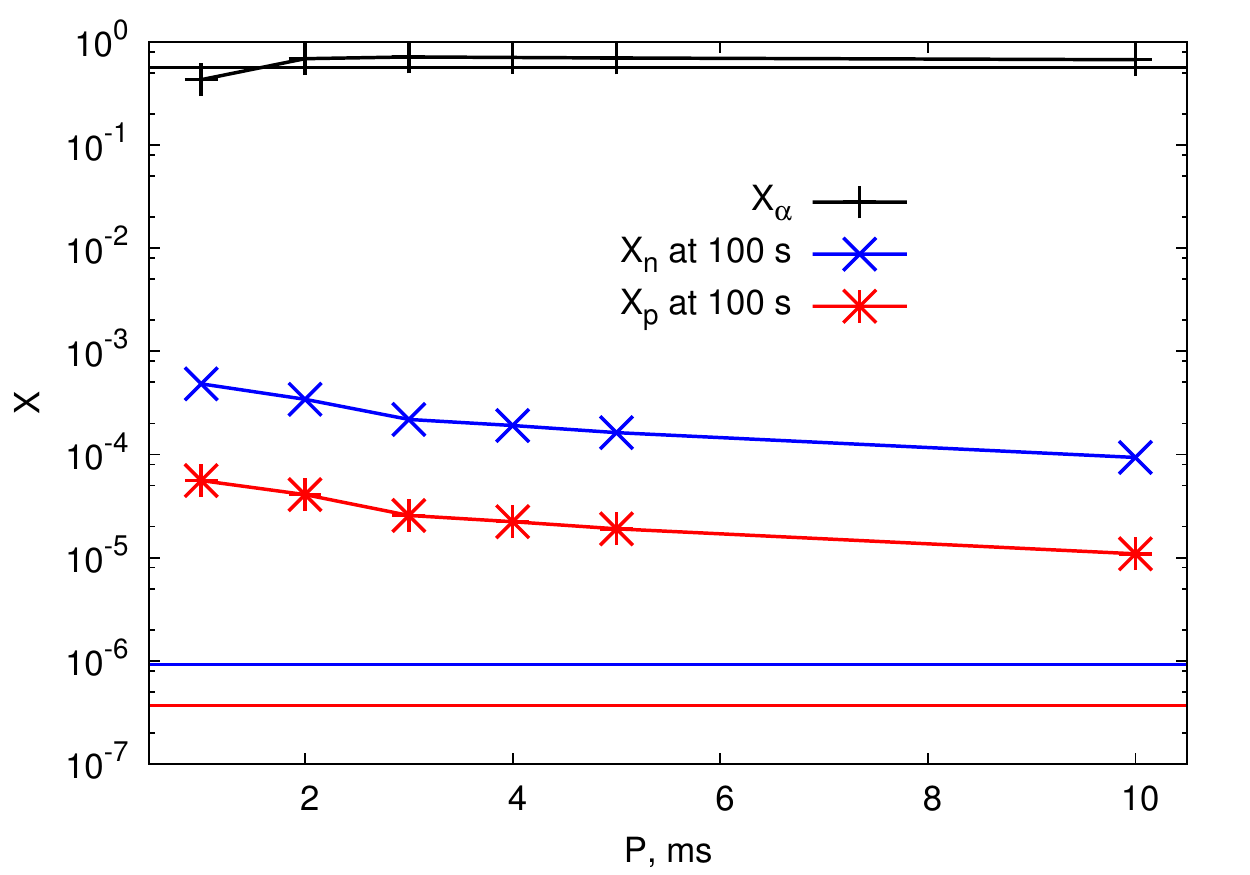}
\caption{Time-averaged mass fractions of $\alpha-$particles, free neutrons and free protons in proto-magnetar winds. We report the mass fractions of neutrons and protons at $t=100$ s, since at that time free neutrons have not yet decayed to protons. The horizontal lines of corresponding colors show these quantities in the spherical model. In spherical wind, the expansion is slower and hence neutrons have more time to capture onto seed nuclei during the $r$-process. This results in much lower value of $X_n$ for spherical winds.}
\label{fig:xnxpxa}
\end{figure}

The birth of millisecond magnetars may also give rise to collimated relativistic jets, which in some cases are sufficiently powerful to break out of the progenitor stars and may power gamma-ray bursts (GRB; \citealt{Usov92,Thompson+04,Bucciantini+08,Metzger+11a}a).  The acceleration of relativistic particles due to shocks or magnetic reconnection within GRB jets is also considered a promising source for the origin of ultra-high energy cosmic rays (UHECRs; \citealt{Waxman95}).  If the outflows from millisecond magnetars indeed feed GRB jets, then the composition of the UHECRs accelerated within such a jet should contain a large quantity of heavy nuclei (\citealt{Metzger+11b}b; see also \citealt{Horiuchi+12}).\footnote{This UHECR source from magnetars is notably distinct from scenarios which invoke particle acceleration within the hot nebula inflated by the magnetar wind behind the supernova ejecta in cases when a GRB jet does not escape from the star (\citealt{Arons03,Fang+12,Lemoine+15,Piro&Kollmeier16}).}  

For rotation periods of $P \sim 1-2$ ms, the ratio of the Poynting flux luminosity to the baryon loading of the magnetar wind is sufficiently high that the resulting jet could reach an asymptotic bulk Lorentz factor of  $\Gamma\approx 100-1000$ (e.g.\citealt{Thompson+04}, \citealt{Metzger+07}).  \citet{Metzger+11b} show that this is sufficient to produce the observed gamma-ray emission and to accelerate nuclei within the jet to ultra-high energies $\gtrsim 10^{19}$ eV per particle.  GRBs associated with magnetar birth therefore provides a natural explanation for the otherwise puzzling observation by the Pierre Auger Observatory that the highest energies UHECRs are composed of heavy nuclei instead of protons (\citealt{Abraham+10}, see however \citealt{Abbasi+05}).  
This model is also consistent with constraints on the non-detection of high energy neutrinos coincident with GRBs with IceCube (\citealt{Abbasi+11}), since nuclei typically lose energy through other processes than photo-pion production, and hence are not expected to be accompanied by a neutrino flux as large as that predicted for proton-dominated compositions.

Figure \ref{fig:xnxpxa} shows the composition of proto-magnetar jets is predicted to be roughly $X_{\rm He} \approx 0.6$ by mass in helium, with the remainder $X_{\rm h} \simeq 1-X_{\rm He} \approx 0.4$ in heavy nuclei of average mass $\bar{A}\approx 95$ (Fig.~\ref{fig:abar}, Table \ref{table:summary}).  The latter result is non-trivial, because if the expansion time of the outflow were sufficiently short, then neutron capture reactions could in principle freeze-out with an order unity mass fraction of free neutrons.  However, we find that the free neutron mass fraction at 100 s is always small $X_n \lesssim 10^{-3}$ (Fig.~\ref{fig:xnxpxa}).  Such a low free neutron fraction would disfavor models in which the GRB prompt emission is powered by the relative kinetic energy in the jet between its neutral (neutron) and charged (protons, nuclei) constituents \citep{Beloborodov10}, unless the nuclei are destroyed in the jet by photo-disintegration before the radius at which neutrons collisionally decouple (see below).  

The prediction of UHECRs dominated by nuclei with $A \sim 100$ has important implications for the energy spectrum and pathlength of UHECRs through the intergalactic medium.  The normal Grezin-Zatsepin-Kuzmin (GZK) cut-off in the cosmic ray spectrum occurs due to inelastic pion production by high energy protons which interact with the cosmic microwave background or extragalactic background light (\citealt{Greisen66,Zatsepin&Kuzmin66}).  At a fixed (measured) cosmic ray energy $E$, a nucleus of mass $A$ has a bulk Lorentz-factor which is a factor of $A$ times smaller than a proton of the same energy.  Naively, one would therefore expect the effective GZK cut-off energy of a nucleus (interacting with background radiation of a fixed temperature like the CMB) to be $A$ time larger than that for a proton.  However, nuclei suffer from other loss processes, the most important one being Giant Dipole Resonances (GDR), which once excited cause the nucleus to shed free nucleons or $\alpha-$particles, reducing its energy.

\citet{Metzger+11b} show that a nucleus of initial mass $A=56 A_{56}$, energy $E=10^{20} E_{20}$ eV travels a distance of
\be
 \chi_{75}=170 E_{20}^{-1.5}A_{56}^{1.3} \bar{n}_{\rm ej}^{-1} \quad {\rm Mpc}
\ee 
before losing 25 \% of its initial energy, where $\bar{n}_{\rm ej}^{-1}\approx 1$ is the mean number of nucleons ejected per GDR excitation.  For an iron nucleus, the mean free path with respect to EBL interaction is coincidentally the same as mean free path for protons with respect to CMB pair production, resulting in a similar GZK cut-off to the proton case.  However, $\chi_{75} \propto A^{1.3}$ implies that for $A=95$ nuclei from our models (Fig.~\ref{fig:abar}), the mean free path is 2 times larger than for protons.  Thus we expect the GZK-like cut-off in the cosmic ray energy spectrum to extend farther for proto-magnetar wind composition than the usual cut-off predicted for protons or iron nuclei.  By contrast, the mean free path for helium is much shorter than iron, such the composition arriving at Earth is expected to be dominated by heavy nuclei at the highest energies.    As already mentioned, the highest energy UHECRs do appear to be heavier than protons \citep{Abraham+10}, although the precise composition is model- and calibration-dependent and hence remains uncertain.  

\section{Conclusions}
We have explored the nucleosynthetic yield of the millisecond magnetar birth, using wind trajectories based on the force-free magnetosphere geometry from V14 (Figs.~\ref{fig:schematic}, \ref{fig:approxcube}) and using a parameterized model for the PNS neutrino cooling evolution from \citet{Roberts+12b} (Fig.~\ref{fig:yelnut}).  Our main conclusions are summarized as follows:
\begin{itemize}
\item{Neutrino-heated winds from millisecond magnetars with rotation periods $P \sim 1-10$ ms produce heavy element abundance distributions that extend to higher atomic number than that from otherwise equivalent spherical winds with the same $Y_e(t)$ (Fig.~\ref{fig:elmej}).  This increase in the neutron-to-seed ratio is driven mainly by the faster expansion rate in proto-magnetar winds caused by the faster divergence of the area function and due to additional magneto-centrifugal acceleration (V14).  For the fastest rotation periods, $P = 1-2$ ms, it is also driven by the lower value of $Y_e$ resulting from centrifugal acceleration.

Unfortunately, a direct detailed comparison of our predicted abundances to data (e.g.~in the solar system or on metal-poor stars) is hindered by the larger uncertainties in the time evolution of $Y_e$ in the wind during the PNS cooling phase.}
\item{The heaviest elements are synthesized by outflows emerging along flux tubes with latitude $\theta \approx \theta_{\rm max}$, i.e. those in outflows which graze the closed zone and pass near the equatorial plane outside the light cylinder (Fig.~\ref{fig:compnuc}).  These fields lines also dominate the total mass budget of the wind due to their larger fraction of the total solid angle and, to a lesser degree, enhancements in the mass loss rate per unit area due to centrifugal acceleration.} 
\item{The total ejecta mass for $P=2-5$ ms is greatly reduced, by a factor of $\gtrsim 10$, due to the small fraction of stellar surface threaded by the open magnetic flux, as compared to a spherical wind (\citealt{Thompson03}). Such suppression in the outflow fraction could be a generic feature of PNS winds if strong, dynamically-important magnetic fields are transient but ubiquitous, e.g.~as would be the case if they are present only during the convective phases of the PNS cooling evolution. Such uncertainties, in addition to the uncertain $Y_e$ evolution make it challenging to place constraints on the birth rate of magnetars based on their nucleosynthesis abundances (Fig.~\ref{fig:rates}).  For the most rapidly rotating case $P=1$ ms, the total mass loss is instead enhanced by a factor of 3 over the spherical case due to additional magneto-centrifugal acceleration (Table \ref{table:summary}).}
\item{Due to dependence of the charged particle process and weak $r$-process pattern on the magnetic field strength and rotation rate of PNSs, natural variations in these quantities between different core collapse events could contribute to the diversity of abundances observed on metal-poor stars (Fig.~\ref{fig:hd}).  This $r$-process site in the PNS phase (post-explosion) is notably distinct from that discussed in the context of MHD supernovae (e.g.~\citealt{Winteler+12}).  The latter, which invoke low-$Y_e$ ejecta, have the potential to produce a greater total $r$-process yield; however, such extreme magnetars are likely rarer than those discussed here due to the larger angular momentum of the stellar progenitor core required to produce a maximally-spinning PNS versus the slower $\sim 2-5$ ms periods described here. 

Additional diversity in the $r$-process abundances of proto-magnetar winds, not considered in detail here, could result from variations in the magnetic inclination angle.  Magnetic dissipation within the resulting striped wind could result in additional wind heating and concomitant entropy gain prior to seed formation, facilitating a heavier $r$-process than the aligned case focused on here.}
\item{If proto-magnetars are the central engines of GRBs, their relativistic jets should contain an order unity mass fraction of heavy nuclei with $\bar{A}\approx 100$.  Subsequent particle acceleration in such a jet could produce UHECRs with a heavy composition (\citealt{Abraham+10}) and an energy spectrum that extends roughly a factor of 2 above the nominal GZK cut-off for protons or iron nuclei (\citealt{Egorova+04}).  Better statistics and modeling of the UHECR energy spectrum, as well as a more firm measurement of the composition-dependent UHECR spectrum, is needed to test this prediction and its subtle differences with the normal proton-dominated model predictions.}
\end{itemize}

\section*{Acknowledgements} 
\label{lastpage}
ADV and BDM gratefully acknowledge support from the National Science Foundation (AST-1410950, AST-1615084), NASA through the Astrophysics Theory Program (NNX16AB30G) and the Fermi Guest Investigator Program (NNX15AU77G, NNX16AR73G), the Research Corporation for Science Advancement Scialog Program (RCSA 23810), and the Alfred P.~Sloan Foundation.


\begin{thebibliography}{}

\bibitem[\protect\citeauthoryear{{Abbasi}, {Abdou}, {Abu-Zayyad}, {Adams},
  {Aguilar}, {Ahlers}, {Andeen}, {Auffenberg}, {Bai}, {Baker} \& et
  al.}{{Abbasi} et~al.}{2011}]{Abbasi+11}
{Abbasi} R.,  {Abdou} Y.,  {Abu-Zayyad} T.,  {Adams} J.,  {Aguilar} J.~A.,
  {Ahlers} M.,  {Andeen} K.,  {Auffenberg} J.,  {Bai} X.,  {Baker} M.,    et
  al. 2011, Physical Review Letters, 106, 141101

\bibitem[\protect\citeauthoryear{{Abbasi}, {Abu-Zayyad}, {Archbold}, {Atkins},
  {Bellido}, {Belov}, {Belz}, {BenZvi}, {Bergman}, {Boyer}, {Burt}, {Cao} \&
  {High Resolution Fly's Eye Collaboration}}{{Abbasi} et~al.}{2005}]{Abbasi+05}
{Abbasi} R.~U.,  {Abu-Zayyad} T.,  {Archbold} G.,  {Atkins} R.,  {Bellido} J.,
  {Belov} K.,  {Belz} J.~W.,  {BenZvi} S.,  {Bergman} D.~R.,  {Boyer} J.,
  {Burt} G.~W.,  {Cao} Z.,    {High Resolution Fly's Eye Collaboration} 2005,
  \apj, 622, 910

\bibitem[\protect\citeauthoryear{{Abraham}, {Abreu}, {Aglietta}, {Ahn},
  {Allard}, {Allekotte}, {Allen}, {Alvarez-Mu{\~n}iz}, {Ambrosio},
  {Anchordoqui} \& et al.}{{Abraham} et~al.}{2010}]{Abraham+10}
{Abraham} J.,  {Abreu} P.,  {Aglietta} M.,  {Ahn} E.~J.,  {Allard} D.,
  {Allekotte} I.,  {Allen} J.,  {Alvarez-Mu{\~n}iz} J.,  {Ambrosio} M.,
  {Anchordoqui} L.,    et al. 2010, Physical Review Letters, 104, 091101

\bibitem[\protect\citeauthoryear{{Arcones}, {Janka} \& {Scheck}}{{Arcones}
  et~al.}{2007}]{Arcones+07}
{Arcones} A.,  {Janka} H.-T.,    {Scheck} L.,  2007, \aap, 467, 1227

\bibitem[\protect\citeauthoryear{{Arcones} \& {Montes}}{{Arcones} \&
  {Montes}}{2011}]{Arcones&Montes11}
{Arcones} A.,  {Montes} F.,  2011, \apj, 731, 5

\bibitem[\protect\citeauthoryear{{Arnould}, {Goriely} \& {Takahashi}}{{Arnould}
  et~al.}{2007}]{Arnould+07}
{Arnould} M.,  {Goriely} S.,    {Takahashi} K.,  2007, \physrep, 450, 97

\bibitem[\protect\citeauthoryear{{Arons}}{{Arons}}{2003}]{Arons03}
{Arons} J.,  2003, \apj, 589, 871

\bibitem[\protect\citeauthoryear{{Banerjee}, {Haxton} \& {Qian}}{{Banerjee}
  et~al.}{2011}]{Banerjee+11}
{Banerjee} P.,  {Haxton} W.~C.,    {Qian} Y.-Z.,  2011, Physical Review
  Letters, 106, 201104

\bibitem[\protect\citeauthoryear{{Beloborodov}}{{Beloborodov}}{2010}]{Beloborodov10}
{Beloborodov} A.~M.,  2010, \mnras, 407, 1033

\bibitem[\protect\citeauthoryear{{Bucciantini}, {Quataert}, {Arons}, {Metzger}
  \& {Thompson}}{{Bucciantini} et~al.}{2008}]{Bucciantini+08}
{Bucciantini} N.,  {Quataert} E.,  {Arons} J.,  {Metzger} B.~D.,    {Thompson}
  T.~A.,  2008, \mnras, 383, L25

\bibitem[\protect\citeauthoryear{{Burbidge}, {Burbidge}, {Fowler} \&
  {Hoyle}}{{Burbidge} et~al.}{1957}]{Burbidge+57}
{Burbidge} E.~M.,  {Burbidge} G.~R.,  {Fowler} W.~A.,    {Hoyle} F.,  1957,
  Reviews of Modern Physics, 29, 547

\bibitem[\protect\citeauthoryear{{Cameron}}{{Cameron}}{1957}]{Cameron57}
{Cameron} A.~G.~W.,  1957, \aj, 62, 9

\bibitem[\protect\citeauthoryear{{Cardall} \& {Fuller}}{{Cardall} \&
  {Fuller}}{1997}]{Cardall&Fuller97}
{Cardall} C.~Y.,  {Fuller} G.~M.,  1997, \apjl, 486, L111

\bibitem[\protect\citeauthoryear{{Cerutti} \& {Beloborodov}}{{Cerutti} \&
  {Beloborodov}}{2016}]{Cerutti&Beloborodov16}
{Cerutti} B.,  {Beloborodov} A.~M.,  2016, \ssr

\bibitem[\protect\citeauthoryear{Cyburt, Amthor, Ferguson, Meisel, Smith,
  Warren, Heger, Hoffman, Rauscher, Sakharuk, Schatz, Thielemann \&
  Wiescher}{Cyburt et~al.}{2010}]{cyburt:10}
Cyburt R.~H.,  Amthor A.~M.,  Ferguson R.,  Meisel Z.,  Smith K.,  Warren S.,
  Heger A.,  Hoffman R.~D.,  Rauscher T.,  Sakharuk A.,  Schatz H.,  Thielemann
  F.~K.,    Wiescher M.,  2010, \apjs, 189, 240

\bibitem[\protect\citeauthoryear{{Duan} \& {Qian}}{{Duan} \&
  {Qian}}{2004}]{Duan&Qian04}
{Duan} H.,  {Qian} Y.,  2004, Phs.~Rev.~D, 69, 123004

\bibitem[\protect\citeauthoryear{{Duncan}, {Shapiro} \& {Wasserman}}{{Duncan}
  et~al.}{1986}]{Duncan+86}
{Duncan} R.~C.,  {Shapiro} S.~L.,    {Wasserman} I.,  1986, \apj, 309, 141

\bibitem[\protect\citeauthoryear{{Egorova}, {Glushkov}, {Ivanov}, {Knurenko},
  {Kolosov}, {Krasilnikov}, {Makarov}, {Mikhailov}, {Olzoev}, {Pravdin},
  {Sabourov}, {Sleptsov} \& {Struchkov}}{{Egorova} et~al.}{2004}]{Egorova+04}
{Egorova} V.~P.,  {Glushkov} A.~V.,  {Ivanov} A.~A.,  {Knurenko} S.~P.,
  {Kolosov} V.~A.,  {Krasilnikov} A.~D.,  {Makarov} I.~T.,  {Mikhailov} A.~A.,
  {Olzoev} V.~V.,  {Pravdin} M.~I.,  {Sabourov} A.~V.,  {Sleptsov} I.~Y.,
  {Struchkov} G.~G.,  2004, Nuclear Physics B Proceedings Supplements, 136, 3

\bibitem[\protect\citeauthoryear{{Eichler}, {Livio}, {Piran} \&
  {Schramm}}{{Eichler} et~al.}{1989}]{Eichler+89}
{Eichler} D.,  {Livio} M.,  {Piran} T.,    {Schramm} D.~N.,  1989, Nature, 340,
  126

\bibitem[\protect\citeauthoryear{{Fang}, {Kotera} \& {Olinto}}{{Fang}
  et~al.}{2012}]{Fang+12}
{Fang} K.,  {Kotera} K.,    {Olinto} A.~V.,  2012, \apj, 750, 118

\bibitem[\protect\citeauthoryear{{Fern{\'a}ndez} \& {Metzger}}{{Fern{\'a}ndez}
  \& {Metzger}}{2013}]{Fernandez&Metzger13}
{Fern{\'a}ndez} R.,  {Metzger} B.~D.,  2013, \mnras, 435, 502

\bibitem[\protect\citeauthoryear{{Fischer}, {Mart{\'{\i}}nez-Pinedo}, {Hempel}
  \& {Liebend{\"o}rfer}}{{Fischer} et~al.}{2012}]{Fischer+12}
{Fischer} T.,  {Mart{\'{\i}}nez-Pinedo} G.,  {Hempel} M.,    {Liebend{\"o}rfer}
  M.,  2012, \prd, 85, 083003

\bibitem[\protect\citeauthoryear{{Frankel} \& {Metropolis}}{{Frankel} \&
  {Metropolis}}{1947}]{frankel:47}
{Frankel} S.,  {Metropolis} N.,  1947, Physical Review, 72, 914

\bibitem[\protect\citeauthoryear{{Freiburghaus}, {Rosswog} \&
  {Thielemann}}{{Freiburghaus} et~al.}{1999}]{Freiburghaus+99}
{Freiburghaus} C.,  {Rosswog} S.,    {Thielemann} F.,  1999, ApJ, 525, L121

\bibitem[\protect\citeauthoryear{{Fr{\"o}hlich}, {Hauser}, {Liebend{\"o}rfer},
  {Mart{\'{\i}}nez-Pinedo}, {Thielemann}, {Bravo}, {Zinner}, {Hix}, {Langanke},
  {Mezzacappa} \& {Nomoto}}{{Fr{\"o}hlich} et~al.}{2006}]{Frohlich+06}
{Fr{\"o}hlich} C.,  {Hauser} P.,  {Liebend{\"o}rfer} M.,
  {Mart{\'{\i}}nez-Pinedo} G.,  {Thielemann} F.-K.,  {Bravo} E.,  {Zinner}
  N.~T.,  {Hix} W.~R.,  {Langanke} K.,  {Mezzacappa} A.,    {Nomoto} K.,  2006,
  \apj, 637, 415

\bibitem[\protect\citeauthoryear{{Fuller}, {Fowler} \& {Newman}}{{Fuller}
  et~al.}{1982}]{fuller:82}
{Fuller} G.~M.,  {Fowler} W.~A.,    {Newman} M.~J.,  1982, \apjs, 48, 279

\bibitem[\protect\citeauthoryear{{Gaensler} \& {Slane}}{{Gaensler} \&
  {Slane}}{2006}]{Gaensler&Slane06}
{Gaensler} B.~M.,  {Slane} P.~O.,  2006, \araa, 44, 17

\bibitem[\protect\citeauthoryear{{Goriely}, {Bauswein} \& {Janka}}{{Goriely}
  et~al.}{2011}]{Goriely+11}
{Goriely} S.,  {Bauswein} A.,    {Janka} H.-T.,  2011, \apjl, 738, L32

\bibitem[\protect\citeauthoryear{{Goriely}, {Bauswein}, {Just}, {Pllumbi} \&
  {Janka}}{{Goriely} et~al.}{2015}]{Goriely+15}
{Goriely} S.,  {Bauswein} A.,  {Just} O.,  {Pllumbi} E.,    {Janka} H.-T.,
  2015, \mnras, 452, 3894

\bibitem[\protect\citeauthoryear{Greisen}{Greisen}{1966}]{Greisen66}
Greisen K.,  1966, Phys. Rev. Lett., 16, 748

\bibitem[\protect\citeauthoryear{{Guilet} \& {M{\"u}ller}}{{Guilet} \&
  {M{\"u}ller}}{2015}]{Guilet&Muller15}
{Guilet} J.,  {M{\"u}ller} E.,  2015, \mnras, 450, 2153

\bibitem[\protect\citeauthoryear{{Hoffman}, {Woosley} \& {Qian}}{{Hoffman}
  et~al.}{1997}]{Hoffman+97}
{Hoffman} R.~D.,  {Woosley} S.~E.,    {Qian} Y.-Z.,  1997, \apj, 482, 951

\bibitem[\protect\citeauthoryear{{Honda}, {Aoki}, {Ishimaru} \&
  {Wanajo}}{{Honda} et~al.}{2007}]{Honda+07}
{Honda} S.,  {Aoki} W.,  {Ishimaru} Y.,    {Wanajo} S.,  2007, \apj, 666, 1189

\bibitem[\protect\citeauthoryear{{Honda}, {Aoki}, {Ishimaru}, {Wanajo} \&
  {Ryan}}{{Honda} et~al.}{2006}]{Honda+06}
{Honda} S.,  {Aoki} W.,  {Ishimaru} Y.,  {Wanajo} S.,    {Ryan} S.~G.,  2006,
  \apj, 643, 1180

\bibitem[\protect\citeauthoryear{{Horiuchi}, {Murase}, {Ioka} \&
  {M{\'e}sz{\'a}ros}}{{Horiuchi} et~al.}{2012}]{Horiuchi+12}
{Horiuchi} S.,  {Murase} K.,  {Ioka} K.,    {M{\'e}sz{\'a}ros} P.,  2012, \apj,
  753, 69

\bibitem[\protect\citeauthoryear{{Hotokezaka}, {Piran} \& {Paul}}{{Hotokezaka}
  et~al.}{2015}]{Hotokezaka+15}
{Hotokezaka} K.,  {Piran} T.,    {Paul} M.,  2015, Nature Physics, 11, 1042

\bibitem[\protect\citeauthoryear{{Ji}, {Frebel}, {Chiti} \& {Simon}}{{Ji}
  et~al.}{2016}]{Ji+16}
{Ji} A.~P.,  {Frebel} A.,  {Chiti} A.,    {Simon} J.~D.,  2016, \nat, 531, 610

\bibitem[\protect\citeauthoryear{{Just}, {Bauswein}, {Pulpillo}, {Goriely} \&
  {Janka}}{{Just} et~al.}{2015}]{Just+15}
{Just} O.,  {Bauswein} A.,  {Pulpillo} R.~A.,  {Goriely} S.,    {Janka} H.-T.,
  2015, \mnras, 448, 541

\bibitem[\protect\citeauthoryear{{Kajino}, {Otsuki}, {Wanajo}, {Orito} \&
  {Mathews}}{{Kajino} et~al.}{2000}]{Kajino+00}
{Kajino} T.,  {Otsuki} K.,  {Wanajo} S.,  {Orito} M.,    {Mathews} G.~J.,
  2000, p.~80

\bibitem[\protect\citeauthoryear{{Kasen} \& {Bildsten}}{{Kasen} \&
  {Bildsten}}{2010}]{Kasen&Bildsten10}
{Kasen} D.,  {Bildsten} L.,  2010, \apj, 717, 245

\bibitem[\protect\citeauthoryear{{Lai} \& {Qian}}{{Lai} \&
  {Qian}}{1998}]{Lai&Qian98}
{Lai} D.,  {Qian} Y.-Z.,  1998, \apj, 505, 844

\bibitem[\protect\citeauthoryear{{Langanke} \&
  {Mart{\'{\i}}nez-Pinedo}}{{Langanke} \&
  {Mart{\'{\i}}nez-Pinedo}}{2000}]{langanke:00}
{Langanke} K.,  {Mart{\'{\i}}nez-Pinedo} G.,  2000, Nucl. Phys. A, 673, 481

\bibitem[\protect\citeauthoryear{{Lattimer} \& {Schramm}}{{Lattimer} \&
  {Schramm}}{1974}]{Lattimer&Schramm74}
{Lattimer} J.~M.,  {Schramm} D.~N.,  1974, \apjl, 192, L145

\bibitem[\protect\citeauthoryear{{Lemoine}, {Kotera} \& {P{\'e}tri}}{{Lemoine}
  et~al.}{2015}]{Lemoine+15}
{Lemoine} M.,  {Kotera} K.,    {P{\'e}tri} J.,  2015, \jcap, 7, 016

\bibitem[\protect\citeauthoryear{{Lippuner} \& {Roberts}}{{Lippuner} \&
  {Roberts}}{2015}]{Lippuner&Roberts15}
{Lippuner} J.,  {Roberts} L.~F.,  2015, \apj, 815, 82

\bibitem[\protect\citeauthoryear{{Lyubarsky} \& {Kirk}}{{Lyubarsky} \&
  {Kirk}}{2001}]{Lyubarsky&Kirk01}
{Lyubarsky} Y.,  {Kirk} J.~G.,  2001, \apj, 547, 437

\bibitem[\protect\citeauthoryear{{Macias} \& {Ramirez-Ruiz}}{{Macias} \&
  {Ramirez-Ruiz}}{2016}]{Macias&Ramirez-Ruiz16}
{Macias} P.,  {Ramirez-Ruiz} E.,  2016, ArXiv e-prints

\bibitem[\protect\citeauthoryear{{Mamdouh}, {Pearson}, {Rayet} \&
  {Tondeur}}{{Mamdouh} et~al.}{2001}]{mamdouh:01}
{Mamdouh} A.,  {Pearson} J.~M.,  {Rayet} M.,    {Tondeur} F.,  2001, Nuc. Phys.
  A, 679, 337

\bibitem[\protect\citeauthoryear{{Martin}, {Perego}, {Arcones}, {Korobkin} \&
  {Thielemann}}{{Martin} et~al.}{2015}]{Martin+15b}
{Martin} D.,  {Perego} A.,  {Arcones} A.,  {Korobkin} O.,    {Thielemann}
  F.-K.,  2015, ArXiv e-prints

\bibitem[\protect\citeauthoryear{{Mart{\'{\i}}nez-Pinedo}, {Fischer}, {Lohs} \&
  {Huther}}{{Mart{\'{\i}}nez-Pinedo} et~al.}{2012}]{MartinezPinedo+12}
{Mart{\'{\i}}nez-Pinedo} G.,  {Fischer} T.,  {Lohs} A.,    {Huther} L.,  2012,
  Physical Review Letters, 109, 251104

\bibitem[\protect\citeauthoryear{{Mathews}, {Bazan} \& {Cowan}}{{Mathews}
  et~al.}{1992}]{Mathews+92}
{Mathews} G.~J.,  {Bazan} G.,    {Cowan} J.~J.,  1992, \apj, 391, 719

\bibitem[\protect\citeauthoryear{{Metzger}, {Giannios} \& {Horiuchi}}{{Metzger}
  et~al.}{2011}]{Metzger+11b}
{Metzger} B.~D.,  {Giannios} D.,    {Horiuchi} S.,  2011, \mnras, 415, 2495

\bibitem[\protect\citeauthoryear{{Metzger}, {Giannios}, {Thompson},
  {Bucciantini} \& {Quataert}}{{Metzger} et~al.}{2011}]{Metzger+11a}
{Metzger} B.~D.,  {Giannios} D.,  {Thompson} T.~A.,  {Bucciantini} N.,
  {Quataert} E.,  2011, \mnras, 413, 2031

\bibitem[\protect\citeauthoryear{{Metzger}, {Quataert} \& {Thompson}}{{Metzger}
  et~al.}{2008}]{Metzger+08a}
{Metzger} B.~D.,  {Quataert} E.,    {Thompson} T.~A.,  2008, \mnras, 385, 1455

\bibitem[\protect\citeauthoryear{{Metzger}, {Thompson} \& {Quataert}}{{Metzger}
  et~al.}{2007}]{Metzger+07}
{Metzger} B.~D.,  {Thompson} T.~A.,    {Quataert} E.,  2007, \apj, 659, 561

\bibitem[\protect\citeauthoryear{{Metzger}, {Thompson} \& {Quataert}}{{Metzger}
  et~al.}{2008}]{Metzger+08b}
{Metzger} B.~D.,  {Thompson} T.~A.,    {Quataert} E.,  2008, \apj, 676, 1130

\bibitem[\protect\citeauthoryear{{Metzger}, {Vurm}, {Hasco{\"e}t} \&
  {Beloborodov}}{{Metzger} et~al.}{2014}]{Metzger+14}
{Metzger} B.~D.,  {Vurm} I.,  {Hasco{\"e}t} R.,    {Beloborodov} A.~M.,  2014,
  \mnras, 437, 703

\bibitem[\protect\citeauthoryear{{Meyer}, {Mathews}, {Howard}, {Woosley} \&
  {Hoffman}}{{Meyer} et~al.}{1992}]{Meyer+92}
{Meyer} B.~S.,  {Mathews} G.~J.,  {Howard} W.~M.,  {Woosley} S.~E.,
  {Hoffman} R.~D.,  1992, \apj, 399, 656

\bibitem[\protect\citeauthoryear{{Moller}, {Sierk}, {Ichikawa} \&
  {Sagawa}}{{Moller} et~al.}{2015}]{moller:15}
{Moller} P.,  {Sierk} A.~J.,  {Ichikawa} T.,    {Sagawa} H.,  2015, ArXiv
  e-prints

\bibitem[\protect\citeauthoryear{{M{\"o}sta}, {Ott}, {Radice}, {Roberts},
  {Schnetter} \& {Haas}}{{M{\"o}sta} et~al.}{2015}]{Mosta+15}
{M{\"o}sta} P.,  {Ott} C.~D.,  {Radice} D.,  {Roberts} L.~F.,  {Schnetter} E.,
    {Haas} R.,  2015, \nat, 528, 376

\bibitem[\protect\citeauthoryear{{M{\"o}sta}, {Richers}, {Ott}, {Haas}, {Piro},
  {Boydstun}, {Abdikamalov}, {Reisswig} \& {Schnetter}}{{M{\"o}sta}
  et~al.}{2014}]{Mosta+14}
{M{\"o}sta} P.,  {Richers} S.,  {Ott} C.~D.,  {Haas} R.,  {Piro} A.~L.,
  {Boydstun} K.,  {Abdikamalov} E.,  {Reisswig} C.,    {Schnetter} E.,  2014,
  \apjl, 785, L29

\bibitem[\protect\citeauthoryear{{Nishimura}, {Sawai}, {Takiwaki}, {Yamada} \&
  {Thielemann}}{{Nishimura} et~al.}{2016}]{Nishimura+16}
{Nishimura} N.,  {Sawai} H.,  {Takiwaki} T.,  {Yamada} S.,    {Thielemann}
  F.-K.,  2016, ArXiv e-prints

\bibitem[\protect\citeauthoryear{{Nishimura}, {Takiwaki} \&
  {Thielemann}}{{Nishimura} et~al.}{2015}]{Nishimura+15}
{Nishimura} N.,  {Takiwaki} T.,    {Thielemann} F.-K.,  2015, \apj, 810, 109

\bibitem[\protect\citeauthoryear{{Oda}, {Hino}, {Muto}, {Takahara} \&
  {Sato}}{{Oda} et~al.}{1994}]{Oda:94}
{Oda} T.,  {Hino} M.,  {Muto} K.,  {Takahara} M.,    {Sato} K.,  1994, Atomic
  Data and Nuclear Data Tables, 56, 231

\bibitem[\protect\citeauthoryear{{Otsuki}, {Tagoshi}, {Kajino} \&
  {Wanajo}}{{Otsuki} et~al.}{2000}]{Otsuki+00}
{Otsuki} K.,  {Tagoshi} H.,  {Kajino} T.,    {Wanajo} S.-y.,  2000, \apj, 533,
  424

\bibitem[\protect\citeauthoryear{{Panov}, {Korneev}, {Rauscher},
  {Mart{\'{\i}}nez-Pinedo}, {Keli{\'c}-Heil}, {Zinner} \& {Thielemann}}{{Panov}
  et~al.}{2010}]{panov:10}
{Panov} I.~V.,  {Korneev} I.~Y.,  {Rauscher} T.,  {Mart{\'{\i}}nez-Pinedo} G.,
  {Keli{\'c}-Heil} A.,  {Zinner} N.~T.,    {Thielemann} F.,  2010, \aap, 513,
  A61

\bibitem[\protect\citeauthoryear{{Perego}, {Rosswog}, {Cabez{\'o}n},
  {Korobkin}, {K{\"a}ppeli}, {Arcones} \& {Liebend{\"o}rfer}}{{Perego}
  et~al.}{2014}]{Perego+14}
{Perego} A.,  {Rosswog} S.,  {Cabez{\'o}n} R.~M.,  {Korobkin} O.,
  {K{\"a}ppeli} R.,  {Arcones} A.,    {Liebend{\"o}rfer} M.,  2014, \mnras,
  443, 3134

\bibitem[\protect\citeauthoryear{{Piro} \& {Kollmeier}}{{Piro} \&
  {Kollmeier}}{2016}]{Piro&Kollmeier16}
{Piro} A.~L.,  {Kollmeier} J.~A.,  2016, \apj, 826, 97

\bibitem[\protect\citeauthoryear{{Podsiadlowski}, {Mazzali}, {Nomoto},
  {Lazzati} \& {Cappellaro}}{{Podsiadlowski} et~al.}{2004}]{Podsiadlowski+04}
{Podsiadlowski} P.,  {Mazzali} P.~A.,  {Nomoto} K.,  {Lazzati} D.,
  {Cappellaro} E.,  2004, \apjl, 607, L17

\bibitem[\protect\citeauthoryear{{Qian} \& {Woosley}}{{Qian} \&
  {Woosley}}{1996}]{Qian&Woosley96}
{Qian} Y.,  {Woosley} S.~E.,  1996, \apj, 471, 331

\bibitem[\protect\citeauthoryear{{Qian}, {Vogel} \& {Wasserburg}}{{Qian}
  et~al.}{1998}]{Qian+98}
{Qian} Y.-Z.,  {Vogel} P.,    {Wasserburg} G.~J.,  1998, \apj, 506, 868

\bibitem[\protect\citeauthoryear{{Qian} \& {Wasserburg}}{{Qian} \&
  {Wasserburg}}{2007}]{Qian&Wasserburg07}
{Qian} Y.-Z.,  {Wasserburg} G.~J.,  2007, \physrep, 442, 237

\bibitem[\protect\citeauthoryear{{Quimby}, {Yuan}, {Akerlof} \&
  {Wheeler}}{{Quimby} et~al.}{2013}]{Quimby+13}
{Quimby} R.~M.,  {Yuan} F.,  {Akerlof} C.,    {Wheeler} J.~C.,  2013, \mnras,
  431, 912

\bibitem[\protect\citeauthoryear{{Ripley}, {Metzger}, {Arcones} \&
  {Mart{\'{\i}}nez-Pinedo}}{{Ripley} et~al.}{2014}]{Ripley+14}
{Ripley} J.~L.,  {Metzger} B.~D.,  {Arcones} A.,    {Mart{\'{\i}}nez-Pinedo}
  G.,  2014, \mnras, 438, 3243

\bibitem[\protect\citeauthoryear{{Roberts}, {Reddy} \& {Shen}}{{Roberts}
  et~al.}{2012}]{Roberts+12b}
{Roberts} L.~F.,  {Reddy} S.,    {Shen} G.,  2012, \prc, 86, 065803

\bibitem[\protect\citeauthoryear{{Roberts}, {Woosley} \& {Hoffman}}{{Roberts}
  et~al.}{2010}]{Roberts+10}
{Roberts} L.~F.,  {Woosley} S.~E.,    {Hoffman} R.~D.,  2010, \apj, 722, 954

\bibitem[\protect\citeauthoryear{{Roederer}}{{Roederer}}{2016}]{Roederer16}
{Roederer} I.~U.,  2016, ArXiv e-prints

\bibitem[\protect\citeauthoryear{{Roederer}, {Cowan}, {Karakas}, {Kratz},
  {Lugaro}, {Simmerer}, {Farouqi} \& {Sneden}}{{Roederer}
  et~al.}{2010}]{Roederer+10}
{Roederer} I.~U.,  {Cowan} J.~J.,  {Karakas} A.~I.,  {Kratz} K.-L.,  {Lugaro}
  M.,  {Simmerer} J.,  {Farouqi} K.,    {Sneden} C.,  2010, \apj, 724, 975

\bibitem[\protect\citeauthoryear{{Sawai} \& {Yamada}}{{Sawai} \&
  {Yamada}}{2016}]{Sawai&Yamada16}
{Sawai} H.,  {Yamada} S.,  2016, \apj, 817, 153

\bibitem[\protect\citeauthoryear{{Sneden}, {Cowan} \& {Gallino}}{{Sneden}
  et~al.}{2008}]{Sneden+08}
{Sneden} C.,  {Cowan} J.~J.,    {Gallino} R.,  2008, \araa, 46, 241

\bibitem[\protect\citeauthoryear{{Spitkovsky}}{{Spitkovsky}}{2006}]{Spitkovsky06}
{Spitkovsky} A.,  2006, \apjl, 648, L51

\bibitem[\protect\citeauthoryear{{Stanek}, {Gnedin}, {Beacom}, {Gould},
  {Johnson}, {Kollmeier}, {Modjaz}, {Pinsonneault}, {Pogge} \&
  {Weinberg}}{{Stanek} et~al.}{2006}]{Stanek+06}
{Stanek} K.~Z.,  {Gnedin} O.~Y.,  {Beacom} J.~F.,  {Gould} A.~P.,  {Johnson}
  J.~A.,  {Kollmeier} J.~A.,  {Modjaz} M.,  {Pinsonneault} M.~H.,  {Pogge} R.,
    {Weinberg} D.~H.,  2006, \actaa, 56, 333

\bibitem[\protect\citeauthoryear{{Sumiyoshi}, {Suzuki}, {Otsuki}, {Terasawa} \&
  {Yamada}}{{Sumiyoshi} et~al.}{2000}]{Sumiyoshi+00}
{Sumiyoshi} K.,  {Suzuki} H.,  {Otsuki} K.,  {Terasawa} M.,    {Yamada} S.,
  2000, \pasj, 52, 601

\bibitem[\protect\citeauthoryear{{Suwa} \& {Tominaga}}{{Suwa} \&
  {Tominaga}}{2015}]{Suwa&Tominaga15}
{Suwa} Y.,  {Tominaga} N.,  2015, \mnras, 451, 282

\bibitem[\protect\citeauthoryear{{Suzuki} \& {Nagataki}}{{Suzuki} \&
  {Nagataki}}{2005}]{Suzuki+Nagataki05}
{Suzuki} T.~K.,  {Nagataki} S.,  2005, \apj, 628, 914

\bibitem[\protect\citeauthoryear{{Takahashi}, {Witti} \& {Janka}}{{Takahashi}
  et~al.}{1994}]{Takahashi+94}
{Takahashi} K.,  {Witti} J.,    {Janka} H.-T.,  1994, \aap, 286, 857

\bibitem[\protect\citeauthoryear{{Takiwaki}, {Kotake} \& {Suwa}}{{Takiwaki}
  et~al.}{2016}]{Takiwaki+16}
{Takiwaki} T.,  {Kotake} K.,    {Suwa} Y.,  2016, \mnras, 461, L112

\bibitem[\protect\citeauthoryear{{Tamborra}, {Raffelt}, {H{\"u}depohl} \&
  {Janka}}{{Tamborra} et~al.}{2012}]{Tamborra+12}
{Tamborra} I.,  {Raffelt} G.~G.,  {H{\"u}depohl} L.,    {Janka} H.-T.,  2012,
  \jcap, 1, 013

\bibitem[\protect\citeauthoryear{{Thielemann}, {Arcones}, {K{\"a}ppeli},
  {Liebend{\"o}rfer}, {Rauscher}, {Winteler}, {Fr{\"o}hlich}, {Dillmann},
  {Fischer}, {Martinez-Pinedo}, {Langanke}, {Farouqi}, {Kratz}, {Panov} \&
  {Korneev}}{{Thielemann} et~al.}{2011}]{Thielemann+11}
{Thielemann} F.-K.,  {Arcones} A.,  {K{\"a}ppeli} R.,  {Liebend{\"o}rfer} M.,
  {Rauscher} T.,  {Winteler} C.,  {Fr{\"o}hlich} C.,  {Dillmann} I.,  {Fischer}
  T.,  {Martinez-Pinedo} G.,  {Langanke} K.,  {Farouqi} K.,  {Kratz} K.-L.,
  {Panov} I.,    {Korneev} I.~K.,  2011, Progress in Particle and Nuclear
  Physics, 66, 346

\bibitem[\protect\citeauthoryear{{Thompson}}{{Thompson}}{2003}]{Thompson03}
{Thompson} T.~A.,  2003, ArXiv Astrophysics e-prints

\bibitem[\protect\citeauthoryear{{Thompson}}{{Thompson}}{2007}]{Thompson07}
{Thompson} T.~A.,  2007, in {Sato} K.,  {Hisano} J.,  eds, Energy Budget in the
  High Energy Universe {Aspects of Neutrino Production in Supernovae}.
pp 251--260

\bibitem[\protect\citeauthoryear{{Thompson}, {Burrows} \& {Meyer}}{{Thompson}
  et~al.}{2001}]{Thompson+01}
{Thompson} T.~A.,  {Burrows} A.,    {Meyer} B.~S.,  2001, \apj, 562, 887

\bibitem[\protect\citeauthoryear{{Thompson}, {Chang} \& {Quataert}}{{Thompson}
  et~al.}{2004}]{Thompson+04}
{Thompson} T.~A.,  {Chang} P.,    {Quataert} E.,  2004, \apj, 611, 380

\bibitem[\protect\citeauthoryear{{Thompson}, {Quataert} \&
  {Burrows}}{{Thompson} et~al.}{2005}]{Thompson+05}
{Thompson} T.~A.,  {Quataert} E.,    {Burrows} A.,  2005, \apj, 620, 861

\bibitem[\protect\citeauthoryear{{Timmes} \& {Swesty}}{{Timmes} \&
  {Swesty}}{2000}]{Timmes&Swesty00}
{Timmes} F.~X.,  {Swesty} F.~D.,  2000, \apjs, 126, 501

\bibitem[\protect\citeauthoryear{{Timokhin}}{{Timokhin}}{2006}]{Timokhin06}
{Timokhin} A.~N.,  2006, \mnras, 368, 1055

\bibitem[\protect\citeauthoryear{{Usov}}{{Usov}}{1992}]{Usov92}
{Usov} V.~V.,  1992, \nat, 357, 472

\bibitem[\protect\citeauthoryear{{Vlasov}, {Metzger} \& {Thompson}}{{Vlasov}
  et~al.}{2014}]{Vlasov+14}
{Vlasov} A.~D.,  {Metzger} B.~D.,    {Thompson} T.~A.,  2014, \mnras, 444, 3537

\bibitem[\protect\citeauthoryear{{Wahl}}{{Wahl}}{2002}]{wahl:02}
{Wahl} A.~C.,  2002, Technical Report LA-13928, Systematics of Fission-Product
  Yields.
Los Alamos National Laboratory, Los Alamos, NM

\bibitem[\protect\citeauthoryear{{Wallner}, {Faestermann}, {Feige},
  {Feldstein}, {Knie}, {Korschinek}, {Kutschera}, {Ofan}, {Paul}, {Quinto},
  {Rugel} \& {Steier}}{{Wallner} et~al.}{2015}]{Wallner+15}
{Wallner} A.,  {Faestermann} T.,  {Feige} J.,  {Feldstein} C.,  {Knie} K.,
  {Korschinek} G.,  {Kutschera} W.,  {Ofan} A.,  {Paul} M.,  {Quinto} F.,
  {Rugel} G.,    {Steier} P.,  2015, Nature Communications, 6, 5956

\bibitem[\protect\citeauthoryear{{Wanajo}, {Sekiguchi}, {Nishimura}, {Kiuchi},
  {Kyutoku} \& {Shibata}}{{Wanajo} et~al.}{2014}]{Wanajo+14}
{Wanajo} S.,  {Sekiguchi} Y.,  {Nishimura} N.,  {Kiuchi} K.,  {Kyutoku} K.,
  {Shibata} M.,  2014, \apjl, 789, L39

\bibitem[\protect\citeauthoryear{{Waxman}}{{Waxman}}{1995}]{Waxman95}
{Waxman} E.,  1995, Physical Review Letters, 75, 386

\bibitem[\protect\citeauthoryear{{Wheeler}, {Yi}, {H{\"o}flich} \&
  {Wang}}{{Wheeler} et~al.}{2000}]{Wheeler+00}
{Wheeler} J.~C.,  {Yi} I.,  {H{\"o}flich} P.,    {Wang} L.,  2000, \apj, 537,
  810

\bibitem[\protect\citeauthoryear{{Winteler}, {K{\"a}ppeli}, {Perego},
  {Arcones}, {Vasset}, {Nishimura}, {Liebend{\"o}rfer} \&
  {Thielemann}}{{Winteler} et~al.}{2012}]{Winteler+12}
{Winteler} C.,  {K{\"a}ppeli} R.,  {Perego} A.,  {Arcones} A.,  {Vasset} N.,
  {Nishimura} N.,  {Liebend{\"o}rfer} M.,    {Thielemann} F.-K.,  2012, \apjl,
  750, L22

\bibitem[\protect\citeauthoryear{{Woods} \& {Thompson}}{{Woods} \&
  {Thompson}}{2006}]{Woods&Thompson06}
{Woods} P.~M.,  {Thompson} C.,  2006, {Soft gamma repeaters and anomalous X-ray
  pulsars: magnetar candidates}.
pp 547--586

\bibitem[\protect\citeauthoryear{{Woosley}}{{Woosley}}{2010}]{Woosley10}
{Woosley} S.~E.,  2010, \apjl, 719, L204

\bibitem[\protect\citeauthoryear{{Woosley} \& {Bloom}}{{Woosley} \&
  {Bloom}}{2006}]{Woosley&Bloom06}
{Woosley} S.~E.,  {Bloom} J.~S.,  2006, ARAA, 44, 507

\bibitem[\protect\citeauthoryear{{Woosley} \& {Hoffman}}{{Woosley} \&
  {Hoffman}}{1992}]{Woosley&Hoffman92}
{Woosley} S.~E.,  {Hoffman} R.~D.,  1992, \apj, 395, 202

\bibitem[\protect\citeauthoryear{{Woosley}, {Wilson}, {Mathews}, {Hoffman} \&
  {Meyer}}{{Woosley} et~al.}{1994}]{Woosley+94}
{Woosley} S.~E.,  {Wilson} J.~R.,  {Mathews} G.~J.,  {Hoffman} R.~D.,
  {Meyer} B.~S.,  1994, \apj, 433, 229

\bibitem[\protect\citeauthoryear{{Wu}, {Fern{\'a}ndez},
  {Mart{\'{\i}}nez-Pinedo} \& {Metzger}}{{Wu} et~al.}{2016}]{Wu+16}
{Wu} M.-R.,  {Fern{\'a}ndez} R.,  {Mart{\'{\i}}nez-Pinedo} G.,    {Metzger}
  B.~D.,  2016, \mnras, 463, 2323

\bibitem[\protect\citeauthoryear{{Wu}, {Fischer}, {Huther},
  {Mart{\'{\i}}nez-Pinedo} \& {Qian}}{{Wu} et~al.}{2014}]{Wu+14}
{Wu} M.-R.,  {Fischer} T.,  {Huther} L.,  {Mart{\'{\i}}nez-Pinedo} G.,
  {Qian} Y.-Z.,  2014, \prd, 89, 061303

\bibitem[\protect\citeauthoryear{Zatsepin \& Kuzmin}{Zatsepin \&
  Kuzmin}{1966}]{Zatsepin&Kuzmin66}
Zatsepin G.~T.,  Kuzmin V.~A.,  1966, JETP letters, 4, 114


\end{thebibliography}

\end{document}